\documentclass[twocolumn,showpacs,aps,prd,superscriptaddress,amsmath,amssymb,floatfix]{revtex4}
\usepackage{graphicx}
\usepackage{bm}

\begin{document}
\title{Asymmetries in ${\bar B}_d^0\to {\bar K}^{*0}\,e^+\,e^-$ decay and contribution of vector resonances}
\author{Alexander Yu. Korchin} \email{korchin@kipt.kharkov.ua}
\affiliation{NSC `Kharkov Institute of Physics and Technology',
61108 Kharkov, Ukraine}
\author{Vladimir A. Kovalchuk}  \email{koval@kipt.kharkov.ua}
\affiliation{NSC `Kharkov Institute of Physics and Technology',
61108 Kharkov, Ukraine}

\date{today}

\begin{abstract}
The fully differential angular distribution for the rare
flavor-changing neutral current decay ${\bar B}_d^0 \to {\bar
K}^{*0} \, (\to K^{-}\, \pi^+) \, e^+\,e^- $ is studied. The
emphasis is placed on accurate treatment of the contribution from
the processes $\bar{B}_d^0 \to \bar{K}^{*0} \, (\to K^{-}\, \pi^+)
\, V $ with intermediate vector resonances $V = \rho(770), \,
\omega(782), \, \phi(1020), \, J/ \psi, \, \psi(2S), \ldots$
decaying into the $e^+ e^-$ pair. The two versions of the
vector-meson-dominance model for the transition $V \, \gamma$ are
used and tested. The present method of including vector resonances
is also compared with the existing in the literature method. The
electron-positron invariant mass dependence of the branching ratio
and various asymmetries is calculated. The branching ratio,
longitudinal polarization fraction of the ${\bar K}^{*0}$ meson,
transverse asymmetry $A_T^{(2)}$ and forward-backward asymmetry
are compared with data from Belle and CDF, and predictions for
experiments at LHCb are made.
\end{abstract}


\maketitle

\setcounter{footnote}{0}

\section{\label{sec:Introduction}Introduction}

The investigation of rare $B$ decays induced by the
flavor-changing neutral current (FCNC) transitions $b\to s$ and
$b\to d$ represents an important test of the standard model (SM)
and its extensions (see \cite{Antonelli:2009} for a review).

Among the rare decays, the process $b\to s \ell^+\ell^-$, where
the virtual photon is converted to the lepton pair, is of
considerable interest. This decay proceeds through a loop
(penguin) diagram, to which high-mass particles introduced in
various extensions to the SM may contribute with sizable
amplitudes. In this decay the angular distributions and lepton
polarizations can probe the chiral structure of the matrix element
\cite{Grossman:2000, Melikhov:1998, Ali:2000, Kruger:2005,
Bobeth:2008, Altmannshofer:2009, Egede:2010} and thereby effects
of the new physics (NP) beyond the SM.

In order to unambiguously measure effects of NP in the observed
process ${\bar B}_d^0 \to {\bar K}^{*0} \, (\to K^{-}\, \pi^+) \,
\ell^+\, \ell^- $ ($l=e, \, \mu$), mediated by $b\to s
\ell^+\ell^-$ decay, one needs to calculate the SM predictions
with a high accuracy. The amplitude in the SM consists of the
short-distance (SD) and long-distance (LD) contributions. The
former are expressed in terms of the Wilson coefficients $C_i$
calculated in perturbative QCD up to a certain order in
$\alpha_s(\mu)$; they carry information on processes at energy
scales $ \sim m_W, \ m_t$. The LD effects describing the
hadronization process are expressed in terms of matrix elements of
several $b \to s$ operators between the initial $B$ and the $K^*$
final state. These hadronic matrix elements are parameterized in
terms of form factors \cite{Ali:2000} that are calculated in
various approaches (see, e.g., \cite{Ball:2005, Defazio:2006}).

The additional LD effects, originating from intermediate vector
resonances $\rho (770)$, $\omega(782)$, $\phi(1020)$,
$J/\psi(1S)$, $\psi(2S)$,$\ldots$, in general, may complicate
theoretical interpretation and make it more model dependent. The
vector resonances modify the amplitude and thus may induce, for
example, the right-handed currents which are absent in the SM.

Present experimental studies~\cite{Babar:2009, Belle:2009,
CDF:2011} of the $ B \to  K^*  \, \ell^+\, \ell^- $ decay aim at
the search of effects of the NP in the whole region of dilepton
invariant mass $m_{ee} \equiv \sqrt{q^2}$ GeV (here $q= q_+ +
q_-$). In these analyses certain cuts are applied in order to
exclude a rather big charmonia contribution.

Recently also the region of small dilepton invariant mass, $m_{ee}
\lesssim 1$ GeV, attracted attention \cite{Grossman:2000}, as
having a potential for searching signatures of the NP. The authors
of \cite{Lefrancois:2009} analyzed the azimuthal angular
distribution in the decay $\bar{B}^0 \to \bar{K}^{*0} \ell^+
\ell^-$ in this region, to test the possibility to measure this
distribution at the LHCb. They have shown the feasibility of
measurements with small systematic uncertainties. In
Ref.~\cite{Korchin:2010} the influence of the low-lying resonances
$\rho (770)$, $\omega(782)$ and $\phi(1020)$ on differential
branching ratio, polarization fraction of the $K^{0*}$ and
transverse asymmetry $A_{\rm T}^{(2)}$ has been studied.

In the present paper we extend calculations of \cite{Korchin:2010}
to the whole region of dilepton invariant mass up to $m_{ee}^{max}
= m_{B} -m_{K^{*}} =4.39 $ GeV. The effective SM Hamiltonian with
the Wilson coefficients in the next-to-next-to-leading order
(NNLO) approximation is applied. The LD effects mediated by the
resonances, {\it i.e.} $\bar{B}^0 \to \bar{K}^{*0} V \to
\bar{K}^{*0} e^+ e^-$ with $V=\rho(770), \ \omega(782), \
\phi(1020), \ J/\psi, \ \psi(2S), \ldots $, are included
explicitly in terms of the helicity amplitudes of the decays
$\bar{B}^0 \to \bar{K}^{*0} V $. The information on the latter is
taken from experiments if available; otherwise it is taken from
theoretical predictions.

The fully differential angular distribution over the three angles
and dilepton invariant mass  for the four-body decay ${\bar B}_d^0
\to K^{-}\, \pi^+ \, e^+\,e^- $ is analyzed. We define a
convenient set of asymmetries which allows one to extract these
asymmetries from the angular distribution once sufficient
statistics is accumulated. These asymmetries may have sensitivity
to various effects of the NP, although in order to see signatures
of these effects, the resonance contribution should be accurately
evaluated.

One of the ingredients in calculation of the resonance
contribution is the transition vertex $V \, \gamma $. This vertex
is conventionally treated in the vector-meson-dominance (VMD)
model. In the present paper we apply two versions of the VMD model
(called subsequently VMD1 and VMD2) which result in rather
different $V \, \gamma $ vertices, in particular, far from the
vector-meson mass shell $q^2 = m_V^2$. Specifically, due to
explicit gauge-invariant construction of the VMD2 Lagrangian the
$V \, \gamma $ transition is suppressed in the region $q^2 \ll
m_V^2$ (for $V = J/\psi, \, \psi(2S), \ldots$). This observation
may be important for estimation of resonance contribution to those
asymmetries, which are small in the SM.

One should mention that the $c \bar{c}$ vector resonances $J/\psi,
\, \psi(2S), \ldots$ have been included earlier in
Refs.~\cite{Ligeti:1996} in the analysis of the $\bar{B}^0 \to
\bar{K}^{*0} \mu^+ \mu^-$ decay. This method of including
resonances has been originally suggested in \cite{Deshpande:1989}.
In order to see sensitivity of observables to the method of
including the $c \bar{c}$ resonances, we perform calculations
using the two methods, and compare the results.

Results of the present calculations are compared with the recent
data from Belle (KEKB) and CDF (Tevatron) experiments for the
differential branching, asymmetry $A_{\rm T}^{(2)}$, longitudinal
polarization fraction of $K^*$ and forward-backward asymmetry.

The paper is organized as follows. In Sec.~\ref{subsec:angle
distribution} the fully differential angular distribution is
discussed. In Section~\ref{subsec:asymmetries} one-dimensional
distributions and definition of asymmetries are defined.
Section~\ref{subsec:transversity} contains expressions for the
transversity amplitudes in framework of the SM. The models of
vector-meson dominance and contributions of vector resonances to
the amplitudes are discussed in Sec.~\ref{subsec:resonances}.
Results for the dependence of observables on the invariant mass
squared are presented in Sec.~\ref{subsec:observables}. In
Sec.~\ref{subsec:two_approaches} we compare two approaches to
inclusion of vector resonances in the amplitudes of the ${\bar
B}_d^0 \to K^{-}\, \pi^+ \, e^+\,e^- $ decay. In
Sec.~\ref{sec:conclusions} we draw conclusions. In
Appendix~\ref{sec:Appendix} some details of the calculation of the
matrix element and the model of the $B \to K^*$ transition form
factors are described. Appendix~\ref{subsec:vector mesons} deals
with calculation of the $\bar{B}^0 \to \bar{K}^{*0} V$ amplitudes
for the off-mass-shell vector meson $V$.

\section{\label{sec:formalism} Angular distributions and
amplitudes for the ${\bar B}_d^0\to {\bar K}^{*0}\,e^+\,e^-$ decay
}

\subsection{ \label{subsec:angle distribution}
Differential decay rate}

\begin{figure}
\centerline{\includegraphics[width=.45\textwidth]{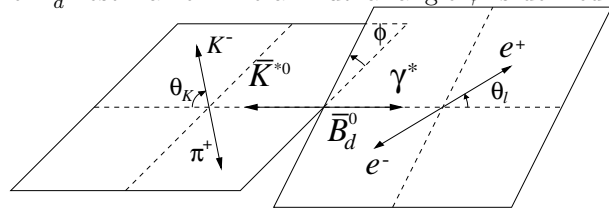}}
\caption{Definition of helicity angles $\theta_l$, $\theta_K$, and
$\phi$, for the decay ${\bar B}_d^0\to {\bar K}^{*0}\,e^+\,e^-$.}
\label{fig1}
\end{figure}

The decay ${\bar B}_d^0\to {\bar K}^{*0}\,e^+\,e^-$, with ${\bar K}^{*0} \to
K^- \pi^+$ on the mass shell~\footnote{This means the narrow-width
approximation for the ${\bar K}^{*0}$ propagator: \ $(k^2 - m_{K^*}^2 +
im_{K^*} \Gamma_{K^*})^{-1} \approx -i \pi \delta(k^2 - m_{K^*}^2) $.}, is
completely described by four independent kinematic variables: the
electron-positron pair invariant-mass squared, $q^2$, and the three angles
$\theta_l$, $\theta_K$, $\phi$. In the helicity frame (Fig.~\ref{fig1}), the
angle $\theta_l\,(\theta_K)$ is defined as the angle between the directions of
motion of $e^+\,(K^-)$ in the $\gamma^*\,({\bar K}^{*0})$ rest frame and the
$\gamma^*\,({\bar K}^{*0})$ in the ${\bar B}_d^0$ rest frame. The azimuthal
angle $\phi$ is defined as the angle between the decay planes of $\gamma^*\to
e^+\,e^-$ and ${\bar K}^{*0} \to K^- \pi^+$ in the ${\bar B}_d^0$ rest frame.
The fully differential angular distribution in these coordinates is given by
\begin{eqnarray}\label{eq:001}
{\cal W}(\hat{q}^2, \theta_l, \theta_K,
\phi)&\equiv&\frac{d^4\,\Gamma}{d\hat{q}^2d\cos\theta_l\,
d\cos\theta_K d\phi}/\frac{d \Gamma}{d\hat{q}^2}\nonumber \\ &=&
\frac{9}{64\,\pi}\sum_{k=1}^{9}\alpha_{k}(q^2)g_{k}(\theta_l,
\theta_K,\phi)\,,
\end{eqnarray}
where the angular terms $g_k$ are defined as
\begin{widetext}
\[g_1=4\sin^2\theta_l\cos^2\theta_K\,,\:
g_2=\left(1+\cos^2\theta_l \right)\sin^2\theta_K\,,\:
g_3=\sin^2\theta_l\sin^2\theta_K\cos2\phi\,,\]
\[g_4=-2\sin^2\theta_l\sin^2\theta_K\sin2\,\phi\,,\:g_5=-\sqrt{2}\sin2\,\theta_l\sin2\,\theta_K\cos\phi\,,
\: g_6=-\sqrt{2}\sin2\,\theta_l\sin2\,\theta_K\sin\phi\,,\]
\[g_7=4\cos\theta_l\sin^2\theta_K\,,\:g_8=-2\sqrt{2}\sin\theta_l\sin2\,\theta_K\cos\phi\,,
\:g_9=-2\sqrt{2}\sin\theta_l\sin2\,\theta_K\sin\phi\,,
\]
and the amplitude terms $\alpha_k$ as
\[\alpha_1=|a_{0}|^2=f_L\,, \: \alpha_2=|a_{\|}|^2+|a_{\perp }|^2=f_{\|}+f_{\perp }\,,\: \alpha_3=|a_{\perp }|^2-|a_{\|}|^2=f_{\perp }-f_{\|}\,,
\: \alpha_4={\rm Im}\left(a_{\|}a_{\perp }^*\right)\,, \:
\alpha_5={\rm Re}\left(a_{0}a_{\|}^*\right)\,,\]
\[\alpha_6={\rm
Im}\left(a_{0}a_{\perp }^*\right)\,,\:\alpha_7={\rm
Re}\left(a_{\|L}a_{\perp L}^*-a_{\|R}a_{\perp R}^*\right)\,,\:
\alpha_8={\rm Re}\left(a_{0L}a_{\perp L}^*-a_{0R}a_{\perp
R}^*\right)\, ,\:\alpha_9={\rm Im}\left(a_{0L}a_{\|
L}^*-a_{0R}a_{\| R}^*\right)\,, \]
\end{widetext} where $\hat{q}^2\equiv q^2/m_B^2$, $m_B$ is the mass of the $B^0_d$ meson, and
\begin{equation}
\frac{d\,\Gamma}{d\hat{q}^2}=m_B\,N^2\hat{q}^2\sqrt{\hat{\lambda}}\left(|A_0|^2+|A_{\|}|^2+|A_{\perp}|^2
\right)  \,. \label{eq:002}
\end{equation}
\begin{equation}
a_i a^*_j \equiv a_{i L}(q^2) a^*_{jL}(q^2)+ a_{iR}(q^2)
a^*_{jR}(q^2)  \, , \label{eq:003}
\end{equation}
\begin{equation}
a_{i L(R)}\equiv \frac{A_{i L(R)}}{\sqrt{\sum_j |A_j|^2}}
\label{eq:004} .
\end{equation}
Here $i,j  = (0, \|, \perp )$, we have neglected the electron mass
$m_e$ and $A_{0L(R)}$, $A_{\|L(R)}$ and $A_{\perp L(R)}$ are the
complex decay amplitudes of the three helicity states in the
transversity basis, $f_L$, $f_{\|}$ and $f_\perp$ are polarization
parameters of the $K^*$ meson, $f_L+f_\|+f_\perp =1$,
$\hat{\lambda}\equiv\lambda(1,\hat{q}^2,
\hat{m}_{K^*}^2)=(1-\hat{q}^2)^2-2(1+\hat{q}^2)\hat{m}_{K^*}^2+\hat{m}_{K^*}^4$,\
$\hat{m}_{K^*}\equiv m_{K^*}/m_B$, where $m_{K^*}$ is the mass of
the $K^{*0}$ meson, and
\[N=|V_{tb}V_{ts}^*|\frac{G_F m_B^2 \alpha_{\rm em}}{32 \,\pi^2 \sqrt{3\,
\pi}}\,.\]
Here, $V_{ij}$ are the Cabibbo-Kobayashi-Maskawa (CKM) matrix
elements \cite{CKM}, $G_F$ is the Fermi coupling constant,
$\alpha_{\rm em}$ is the electromagnetic fine-structure constant.

With its rich multidimensional structure, the differential decay
rate in Eq.~(\ref{eq:001}) has sensitivity to various effects
modifying the SM, such as $CP$ violation beyond the CKM mechanism
and/or right-handed currents. Given sufficient data, all
$\alpha_k$ can, in principle, be completely measured from the full
angular distribution in all three angles $\theta_l$, $\theta_K$,
and $\phi$.

\subsection{ \label{subsec:asymmetries}
One-dimensional angular distributions and asymmetries}

The one-dimensional angular distributions in $\cos\theta_l$ and
$\cos\theta_K$ simply are
\begin{eqnarray}
{\cal W}_{\theta_l}(\hat{q}^2,\cos\theta_l) &\equiv&
\frac{d^2\,\Gamma}{d\hat{q}^2d\cos\theta_l}/\frac{d\,\Gamma}{d\hat{q}^2}=\frac{3}{4}f_L
(1-\cos^2\theta_l) \nonumber \\
&&+\frac{3}{8} (1-f_L)(1+\cos^2\theta_l)\nonumber \\
&&+\frac{d{\bar A}_{\rm FB}^{\rm (l)}}{d\hat{q}^2}\cos\theta_l
 \label{eq:005}
\end{eqnarray}
and
\begin{eqnarray}
{\cal W}_{\theta_K}(\hat{q}^2,\cos\theta_K)&\equiv&
\frac{d^2\,\Gamma}{d\hat{q}^2d\cos\theta_K}/\frac{d\,\Gamma}{d\hat{q}^2}=\frac{3}{2}f_L
\cos^2\theta_K \nonumber \\
&&+\frac{3}{4} (1-f_L)(1-\cos^2\theta_K)
 \label{eq:006},
\end{eqnarray}
where $d{\bar A}_{\rm FB}^{\rm (l)}/d\hat{q}^2$ is the normalized
lepton forward-backward asymmetry
\begin{eqnarray}
\frac{d{\bar A}_{\rm FB}^{\rm
(l)}}{d\hat{q}^2}&\equiv&\int\limits_{-1}^{1}{\rm
sgn}(\cos\theta_l){\cal W}_{\theta_l}(\hat{q}^2,\cos\theta_l)\:d\cos\theta_l\nonumber \\
&=&\frac{3}{2}{\rm Re}(a_{\parallel\,L}\,
a_{\perp\,L}^*-a_{\parallel\,R}\, a_{\perp\,R}^*) \equiv A_7 \, .
\label{eq:007}
\end{eqnarray}
While ${\cal W}_{\theta_K}(\hat{q}^2,\cos\theta_K)$ depends only
on $f_L$, ${\cal W}_{\theta_l}(\hat{q}^2,\cos\theta_l)$ depends
both on $f_L$ and $d{\bar A}_{\rm FB}^{\rm (l)}/d\hat{q}^2$. The
measurement of the lepton forward-backward asymmetry $d{\bar
A}_{\rm FB}^{\rm (l)}/d\hat{q}^2$ alone is not enough to fully
reconstruct the $\cos\theta_l$ distribution. One can then think
about other asymmetries. For any fixed $z$ in the interval
$[-1,1]$, one can define an asymmetry
\begin{equation} \label{eq:008}
{\cal A}_z\equiv \frac{{\cal E}_z-{\cal P}_z}{{\cal E}_z+{\cal
P}_z} ,
\end{equation}
where
\begin{equation}
{\cal E}_z\equiv\int\limits_{-z}^{z}{\cal
W}_{\theta}(\hat{q}^2,\cos\theta)\:d\cos\theta , \label{eq:009}
\end{equation}
and
\begin{equation}
{\cal P}_z\equiv\left(\int\limits_{-1}^{-z}d\cos\theta
+\int\limits_{z}^{1}d\cos\theta \right) {\cal
W}_{\theta}(\hat{q}^2,\cos\theta)\label{eq:010}.
\end{equation}
Measuring the asymmetry ${\cal A}_{z_l}$ for $z=z_l\approx0.596$
($\cos^3\theta_l+3\cos\theta_l-2=0$), we can find the fraction of
the longitudinal polarization of the $K^*$ meson
\begin{equation}
{\cal A}_{z_l}=3(2z_l-1)f_L\approx0.576 f_L \label{eq:011},
\end{equation}
Similarly, measuring the asymmetry ${\cal A}_{z_K}$ for
$z=z_K=2\cos\frac{4\pi}{9}\approx0.347$
($\cos^3\theta_K-3\cos\theta_K+1=0$), we can find the fraction of
the longitudinal polarization of the $K^*$ meson
\begin{equation}
{\cal A}_{z_K}=3(2z_K-1)f_L\approx -0.916 f_L \label{eq:012}.
\end{equation}
Finally, the one-dimensional angular distribution in the angle
$\phi$ between the lepton and meson planes takes the form
\begin{eqnarray} \label{eq:013}
{\cal W}_{\phi}(\hat{q}^2, \phi)
&\equiv&\frac{d^2\,\Gamma}{d\hat{q}^2d\phi}/\frac{d\,\Gamma}{d\hat{q}^2}=\frac{1}{2\pi}
\Bigl(1+\frac{1}{2}\bigl(1\nonumber \\&&-f_L\bigr)A^{(2)}_{\rm
T}\cos2\phi -A_{\rm Im}\sin2\phi\Bigr) ,
\end{eqnarray}
\begin{equation} \label{eq:014}
A^{(2)}_{\rm T}\equiv  \frac{ f_\perp-f_\|}{f_\perp+f_\|}\,, \quad
A_{\rm Im}\equiv {\rm Im}(a_\|a^*_\perp) ,
\end{equation}
where the asymmetry $A^{(2)}_{\rm T}(q^2)$ is sensitive to new
physics from right-handed currents, and the amplitude $A_{\rm
Im}(q^2)$ is sensitive to complex phases in the hadronic matrix
elements. Sometimes $A^{(2)}_{\rm T}(q^2)$ is called transverse
asymmetry~\cite{Kruger:2005}.

Measurement of the angular distribution in the azimuthal angle $\phi$ allows
one to determine the quantities $(1-f_L)A^{(2)}_{\rm T}$ and $A_{\rm Im}$
\begin{eqnarray}
A_3&\equiv&
\Bigl(\int\limits_{0}^{\pi/4}d\,\phi-\int\limits_{\pi/4}^{3\pi/4}d\,\phi
 +\int\limits_{3\pi/4}^{5\pi/4}d\,\phi \nonumber \\
&&-\int\limits_{5\pi/4}^{7\pi/4}d\,\phi
+\int\limits_{7\pi/4}^{2\pi}d\,\phi\Bigr) {\cal
W}_{\phi}(\hat{q}^2, \phi)\nonumber \\
&=& \frac{1}{\pi}(1-f_L)A^{(2)}_{\rm T}=\frac{f_\perp-f_\|}{\pi} ,
\label{eq:015}
\end{eqnarray}
\begin{eqnarray}
A_4&\equiv&\Bigl(\int\limits_{0}^{\pi/2}d\,\phi-\int\limits_{\pi/2}^{\pi}d\,\phi
 +\int\limits_{\pi}^{3\pi/2}d\,\phi
\nonumber \\&&-\int\limits_{3\pi/2}^{2\pi}d\,\phi\Bigr){\cal
W}_{\phi}(\hat{q}^2, \phi) = -\frac{2}{\pi}A_{\rm Im} .
\label{eq:016}
\end{eqnarray}
Measurement of the azimuthal angle dependence of the
forward-backward asymmetry for positrons and $K^-$ mesons
\begin{eqnarray}
&&\frac{d^2{\bar A}_{\rm FB}^{\rm
(Kl)}}{d\hat{q}^2d\phi}\equiv\int\limits_{-1}^{1}{\rm
sgn}(\cos\theta_l)\,d\cos\theta_l \nonumber
\\&\times&\int\limits_{-1}^{1}{\rm
sgn}(\cos\theta_K)\,d\cos\theta_K \, {\cal W}(\hat{q}^2, \theta_l,
\theta_K, \phi)\nonumber \\
 &=&-\frac{\sqrt{2}}{4\pi}\Bigl({\rm
Re}(a_{0}\, a_{\|}^*)\cos\phi +{\rm Im}(a_{0}\,
a_{\perp}^*)\sin\phi\Bigr) ,  \label{eq:017}
\end{eqnarray}
will allow one to find ${\rm Re}(a_{0}\, a_{\|}^*)$ and ${\rm Im}(a_{0}\,
a_{\perp}^*)$
\begin{eqnarray}
A_5&\equiv&\Bigl(\int\limits_{0}^{\pi/2}d\,\phi-\int\limits_{\pi/2}^{3\pi/2}d\,\phi+\int\limits_{3\pi/2}^{2\pi}d\,\phi
 \Bigr)\frac{d^2{\bar A}_{\rm FB}^{\rm
(Kl)}}{d\hat{q}^2d\phi}\nonumber \\ &=& -\frac{\sqrt{2}}{\pi}{\rm
Re}(a_{0}\, a_{\|}^*) , \label{eq:018}
\end{eqnarray}
\begin{equation}
A_6\equiv\Bigl(\int\limits_{0}^{\pi}d\,\phi-\int\limits_{\pi}^{2\pi}d\,\phi
 \Bigr)\frac{d^2{\bar A}_{\rm FB}^{\rm
(Kl)}}{d\hat{q}^2d\phi}= -\frac{\sqrt{2}}{\pi}{\rm Im}(a_{0}\,
a_{\perp}^*) . \label{eq:019}
\end{equation}
Measurement of the azimuthal angle dependence of the
forward-backward asymmetry for $K^-$ mesons
\begin{eqnarray}
&&\frac{d^2{\bar A}_{\rm FB}^{\rm (K)}}{d\hat{q}^2d\phi}\nonumber
\\&\equiv&\int\limits_{-1}^{1}d\cos\theta_l
\int\limits_{-1}^{1}{\rm sgn}(\cos\theta_K)\,d\cos\theta_K \,{\cal
W}(\hat{q}^2, \theta_l,
\theta_K, \phi) \nonumber \\
&=&-\frac{3\sqrt{2}}{16}\Bigl({\rm Re}(a_{0L}\, a_{\perp
L}^*-a_{0R}\, a_{\perp R}^*)\cos\phi \nonumber \\ &&+{\rm Im}(a_{0
L}\, a_{\| L}^*-a_{0 R}\, a_{\| R}^*)\sin\phi\Bigr)
\label{eq:020},
\end{eqnarray}
will allow one to find ${\rm Re}(a_{0L}\, a_{\perp L}^*-a_{0R}\,
a_{\perp R}^*)$ and ${\rm Im}(a_{0 L}\, a_{\| L}^*-a_{0 R}\, a_{\|
R}^*)$
\begin{eqnarray}
A_8&\equiv&\Bigl(\int\limits_{0}^{\pi/2}d\,\phi-\int\limits_{\pi/2}^{3\pi/2}d\,\phi+\int\limits_{3\pi/2}^{2\pi}d\,\phi
 \Bigr)\frac{d^2{\bar A}_{\rm FB}^{\rm
(K)}}{d\hat{q}^2d\phi}\nonumber \\&=& -\frac{3\sqrt{2}}{4}{\rm
Re}(a_{0 L}\, a_{\perp L}^*-a_{0 R}\, a_{\perp R}^*)
,\label{eq:021}
\end{eqnarray}
\begin{eqnarray}
A_9&\equiv&\Bigl(\int\limits_{0}^{\pi}d\,\phi-\int\limits_{\pi}^{2\pi}d\,\phi
 \Bigr)\frac{d^2{\bar A}_{\rm FB}^{\rm
(K)}}{d\hat{q}^2d\phi}\nonumber \\&=& -\frac{3\sqrt{2}}{4}{\rm
Im}(a_{0L}\, a_{\| L}^*-a_{0R}\, a_{\| R}^*) . \label{eq:022}
\end{eqnarray}
%
\subsection{ \label{subsec:transversity}
Transversity amplitudes}

The nonresonant amplitudes follow from the matrix element of the
${\bar B}_d^0 (p)\to {\bar
K}^{*0}(k,\epsilon)\,e^+(q_+)\,e^-(q_-)$ process in
Eq.~(\ref{eq:0A1}),
\begin{eqnarray}
A_{0L,R}^{\rm
NR}&=&\frac{C_0(q^2)}{2\,\hat{m}_{K^*}\sqrt{\hat{q}^2}}\,\Biggl(C_{9V}^{\rm
eff} \mp C_{10A} \nonumber \\ &&+2\hat{m}_b\left(C_{7\gamma}^{\rm
eff}-C_{7\gamma}^{\prime\,\rm
eff}\right)\kappa_0(q^2)\Biggr)\,,\label{eq:023}
\end{eqnarray}
\begin{eqnarray}
A_{\|L,R}^{\rm NR}&=&-\sqrt{2}\,C_{\|}(q^2)\,\Biggl(C_{9V}^{\rm
eff} \mp C_{10A}\nonumber
\\ &&+2\frac{\hat{m}_b}{\hat{q}^2}\left(C_{7\gamma}^{\rm
eff}-C_{7\gamma}^{\prime\,\rm
eff}\right)\kappa_{\|}(q^2)\Biggr)\,,\label{eq:024}
\end{eqnarray}
\begin{eqnarray}
A_{\perp L,R}^{\rm
NR}&=&\sqrt{2\hat{\lambda}}\,C_{\perp}(q^2)\,\Biggl(C_{9V}^{\rm
eff} \mp C_{10A}\nonumber
\\ &&+2\frac{\hat{m}_b}{\hat{q}^2}\left(C_{7\gamma}^{\rm
eff}+C_{7\gamma}^{\prime\,\rm
eff}\right)\kappa_{\perp}(q^2)\Biggr)\,,\label{eq:025}
\end{eqnarray}
where the form factors enter as
\begin{eqnarray}
C_0(q^2)&=&(1-\hat{q}^2-\hat{m}_{K^*}^2)(1+\hat{m}_{K^*})
A_1(q^2)\nonumber \\ &&-\hat{\lambda}
\frac{A_2(q^2)}{1+\hat{m}_{K^*}} \label{eq:026},
\end{eqnarray}
\begin{equation}
C_{\|}(q^2)=(1+\hat{m}_{K^*})A_1(q^2) \label{eq:027},
\end{equation}
\begin{equation}
C_{\perp}(q^2)=\frac{V(q^2)}{1+\hat{m}_{K^*}} \label{eq:028},
\end{equation}
\begin{eqnarray}
\kappa_0(q^2)&\equiv&
\Bigl((1-\hat{q}^2+3\hat{m}_{K^*}^2)(1+\hat{m}_{K^*})T_2(q^2)\nonumber
\\&&-\frac{
\hat{\lambda}}{1-\hat{m}_{K^*}}T_3(q^2)\Bigr)\Bigl((1-\hat{q}^2-\hat{m}_{K^*}^2)\nonumber
\\&&\times(1+\hat{m}_{K^*})^2 A_1(q^2)-\hat{\lambda}\,
A_2(q^2)\Bigr)^{-1} \label{eq:029},
\end{eqnarray}
\begin{equation}
\kappa_{\|}(q^2)\equiv \frac{T_2(q^2)}{A_1(q^2)}(1-\hat{m}_{K^*}),
\label{eq:030}
\end{equation}
\begin{equation}
\kappa_{\perp}(q^2)\equiv \frac{T_1(q^2)}{V(q^2)}(1+\hat{m}_{K^*})
. \label{eq:031}
\end{equation}
In the above formulas the definition $\hat{m}_b\equiv
\overline{m}_b(\mu)/m_B$, $\hat{m}_s\equiv
\overline{m}_s(\mu)/m_B$ are used, and $A_1(q^2), \, A_2(q^2), \,
V(q^2), \, T_1(q^2), \, T_2(q^2), \, T_3(q^2)$ are the $B \to K^*$
transition form factors, specified in Appendix~\ref{sec:Appendix}.

\subsection{\label{subsec:resonances}
 Resonant contribution}

\begin{figure}[tbh]
\centerline{\includegraphics[width=.45\textwidth]{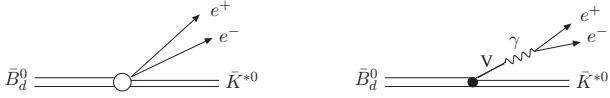}}
\caption{Nonresonant and resonant contributions to the decay
amplitude. } \label{fig:resonances}
\end{figure}

Next, we implement the effects of LD contributions from the decays ${\bar
B}_d^0\to {\bar K}^{*0}\,V$, where $V=\rho^0\,,\omega\,,
\phi\,,J/\psi(1S)\,,\psi(2S)\,,\ldots$ mesons, followed by $V\to e^+\,e^-$ in
the decay ${\bar B}_d^0\to {\bar K}^{*0}\,e^+\,e^-$~(see
Fig.~\ref{fig:resonances}).
We apply vector-meson dominance (VMD) approach. In general, the $V
\, \gamma$ transition can be included into consideration using
various versions of VMD model. In the ``standard'' version (see,
e.g. \cite{Feynman}, chapter~6), the $V \, \gamma $ transition
vertex can be written as
\begin{equation} \langle \gamma (q); \, \mu | V (q); \, \nu\, \rangle = -e f_V Q_V
m_V \, g^{\mu \nu} , \label{eq:032}
\end{equation}
where $q$ is the virtual photon (vector meson) four-momentum,
$g^{\mu \nu}$ is the metric tensor, \  $Q_V$ is the effective
electric charge of the quarks in the vector meson:
\begin{eqnarray}
&&Q_\rho =\frac{1}{\sqrt{2}} \,, \quad  Q_\omega=\frac{1}{3
\sqrt{2}}\,, \quad Q_\phi=- \frac{1}{3}\,,\nonumber \\   &&
Q_{J/\psi} = Q_{\psi(2S)}=\ldots = \frac{2}{3}\,. \label{eq:033}
\end{eqnarray}
The decay constants of neutral vector mesons $f_V$ can be
extracted from their electromagnetic decay width, using
 \begin{equation}
\Gamma_{V \to e^+ e^-} = \frac{4 \pi \alpha_{em}^2 }{3\,m_V} f_V^2
Q_V^2\,. \label{eq:034}
 \end{equation}
This version of VMD model will be called VMD1. The vertex
(\ref{eq:032}) comes from the transition Lagrangian
 \begin{equation}
{\cal L}_{\gamma V} = -e  A^\mu  \sum_{V} f_V Q_V m_V \, V_\mu \,.
\label{eq:035}
 \end{equation}
A more elaborate model (called hereafter VMD2) originates from
Lagrangian
 \begin{equation}
{\cal L}_{\gamma V} = -\frac{e}{2} F^{\mu \nu} \sum_{V} \frac{f_V
Q_V}{m_V} \, V_{\mu \nu}  \label{eq:036}
\end{equation}
where $V_{\mu \nu} \equiv \partial_\mu V_\nu -
\partial_\nu V_\mu$ and $F^{\mu \nu} \equiv \partial^\mu A^\nu -
\partial^\nu A^\mu$ is the electromagnetic field tensor.

Lagrangian (\ref{eq:036}) is explicitly gauge invariant, unlike
Eq.~(\ref{eq:035}), and gives rise to to the $ V \, \gamma^*$
vertex
\begin{equation}
\langle \,\gamma (q); \, \mu | V (q); \, \nu  \,\rangle = -\frac{e
f_V Q_V }{m_V} \, (q^2 g^{\mu \nu} - q^\mu q^\nu )\,,
\label{eq:037}
\end{equation}
This transition vertex is suppressed at small invariant masses,
$q^2 \ll m_V^2$, i.e. in the region far from the vector-meson mass
shell~\footnote{The term $\propto q^\mu q^\nu /q^2 $ in
(\ref{eq:037}) does not contribute when contracted with the
leptonic current}.

Note that these two versions of the VMD model have been discussed
in Refs.~\cite{Klingl:1996by,O'Connell}. The VMD2 version
naturally follows from the Resonance Chiral Theory
\cite{EckerNP321}; in this context VMD2 coupling has been applied
in \cite{Eidelman:2010ta} for studying electron-positron
annihilation into $\pi^0 \pi^0 \gamma$ and $\pi^0 \eta \gamma$
final states.

Parameters of vector resonances are presented in
Table~\ref{tab:param1}.
\begin{table}[th]
\caption{Mass, total width, leptonic decay width and coupling
$f_V$ of vector mesons~\cite{PDG:2010} (experimental uncertainties
are not shown).}
\begin{center}
\begin{tabular}{c c c c c}
\hline \hline $V$ & $m_V{\rm (MeV)}$ &$\Gamma_V{\rm (MeV)}$&$\Gamma_{V\to e^+\,e^-}{\rm (keV)}$&$f_V{\rm (MeV)}$  \\
\hline
$\rho^0$   & $775.49$ & $149.1$ & $7.04$ & $221.2$  \\
$\omega$   &$782.65$  & $8.49$ & $0.60$ & $194.7$   \\
$\phi$     &$1019.455$  & $4.26$ & $1.27$ & $228.6$  \\
$J/\psi$ & $3096.916$ & $0.0929$ & $5.55$ & $416.4$ \\
$\psi(2S)$ & $3686.09$ & $0.304$ & $2.35$ & $295.6$ \\
$\psi(3770)$ & $3772.92$ & $27.3$ & $0.265$ & $100.4$ \\
$\psi(4040)$ & $4039$ & $80$ & $0.86$ & $187.2$ \\
$\psi(4160)$ & $4153$ & $103$ & $0.83$ & $186.5$ \\
$\psi(4415)$ & $4421$ & $62$ & $0.58$ & $160.8$ \\
\hline\hline
\end{tabular}
\end{center}
\label{tab:param1}
\end{table}

Based on VMD approach, we obtain the total amplitude including
nonresonant and resonant parts,
\begin{eqnarray} A_{0
L,R}& = &
\frac{1}{2\,\hat{m}_{K^*}\sqrt{\hat{q}^2}}\Biggl(C_0(q^2)\Bigl(C_{9V}^{\rm
eff} \mp C_{10A} \nonumber \\ &&+2\hat{m}_b\left(C_{7\gamma}^{\rm
eff}-C_{7\gamma}^{\prime\,\rm eff}\right)\kappa_0(q^2)\Bigr)
\nonumber \\
  && + 8\pi^2\sum_{V}C_V D_V^{-1}(\hat{q}^2)
\Bigl(\left(1-\hat{q}^2-\hat{m}_{K^*}^2\right)S_1^V\nonumber \\
&&+\hat{\lambda}\frac{S_2^V}{2}\Bigr)\Biggr)\,, \label{eq:038}
\end{eqnarray}
\begin{eqnarray}
A_{\| L,R}&=&-\sqrt{2}\Biggl(C_{\|}(q^2)\Bigl(C_{9V}^{\rm eff} \mp
C_{10A}\nonumber \\
&&+2\frac{\hat{m}_b}{\hat{q}^2}\left(C_{7\gamma}^{\rm
eff}-C_{7\gamma}^{\prime\,\rm
eff}\right)\kappa_{\|}(q^2)\Bigr)\nonumber \\
&&+8\pi^2\sum_{V}C_V
D_V^{-1}(\hat{q}^2)\,S_1^V\Biggr)\,,\label{eq:039}
\end{eqnarray}
\begin{eqnarray}
A_{\perp
L,R}&=&\sqrt{2\hat{\lambda}}\Biggl(C_{\perp}(q^2)\Bigl(C_{9V}^{\rm
eff} \mp
C_{10A}\nonumber \\
&&+2\frac{\hat{m}_b}{\hat{q}^2}\left(C_{7\gamma}^{\rm
eff}+C_{7\gamma}^{\prime\,\rm
eff}\right)\kappa_{\perp}(q^2)\Bigr)\nonumber \\
&&+4\pi^2\sum_{V}C_V
D_V^{-1}(\hat{q}^2)\,S_3^V\Biggr)\,,\label{eq:040}
\end{eqnarray}
where
\[ D_V(\hat{q}^2) = \hat{q}^2 - \hat{m}_V^2 +
i\hat{m}_V \hat{\Gamma}_V (\hat{q}^2) \] is the usual Breit-Wigner
function for the $V$ meson resonance shape with the
energy-dependent width ${\Gamma}_V ({q}^2)$ \ [$\hat{\Gamma}_V
(\hat{q}^2) = {\Gamma}_V ({q}^2) /m_B $], $\hat{m}_V\equiv
m_V/m_B$, $\hat{\Gamma}_V\equiv \Gamma_V/m_B$, $m_V(\Gamma_V)$ is
the mass (width) of a $V$ meson.
\begin{equation} C_V= \frac{Q_V m_V f_V}{q^2}\,\left(\rm{VMD1}\right)\,,
 \quad C_V= \frac{Q_V f_V}{m_V}\,\left(\rm{VMD2}\right)\, .
\label{eq:041}
\end{equation}

In Eqs.~(\ref{eq:038})-(\ref{eq:040}), $S_i^V$ \ ($i=1,2,3$) are
the invariant amplitudes of the decay $B_d^0\to  K^{*0}\,V$. These
amplitudes are calculated in Appendix~\ref{subsec:vector mesons}.

The energy-dependent widths of light vector resonances $\rho$,
$\omega$ and $\phi$ are chosen as in Ref.~\cite{Korchin:2010}. The
up-dated branching ratios for resonances decays to different
channels are taken from \cite{PDG:2010}. For the $c \bar{c}$
resonances $J/\psi$, \ $\psi (2S)$, $\ldots$ we take the constant
widths.

In order to calculate the resonant contribution to the amplitude
of the ${\bar B}_d^0\to {\bar K}^{*0}\,e^+\,e^-$ decay, one has to
know the amplitudes of the decays ${\bar B}_d^0\to {\bar K}^{*0}\,
\rho$, \ ${\bar B}_d^0\to {\bar K}^{*0}\,\omega, \, {\bar
B}_d^0\to {\bar K}^{*0}\,\phi, \, {\bar B}_d^0\to {\bar
K}^{*0}\,J/\psi, \, {\bar B}_d^0\to {\bar K}^{*0}\,\psi(2S),
\ldots$  At present the amplitudes of the ${\bar B}_d^0\to {\bar
K}^{*0}\,\phi, \, {\bar B}_d^0\to {\bar K}^{*0}\,J/\psi, \, {\bar
B}_d^0\to {\bar K}^{*0}\,\psi(2S) $ decays are known from
experiment~\cite{PDG:2010}, therefore, we use these amplitudes for
calculation of invariant amplitudes in
Appendix~~\ref{subsec:vector mesons} in Table~\ref{tab:ampl}. For
the light resonances $\rho$ and $ \omega$ we use the theoretical
prediction~\cite{Chen:2006} for the decay amplitudes. At the same
time, we are not aware of a similar prediction for the higher $c
\bar{c}$ resonances, such as $\psi(3770)$ an so on, therefore we
do not include contribution of these resonances to amplitudes.

The parameters of the model are indicated in
Table~\ref{tab:param2}. The SM Wilson coefficients have been
obtained in \cite{Altmannshofer:2009} at the scale $\mu=4.8$ GeV
to NNLO accuracy. In our notation (see Appendix~\ref{subsec:matrix
element}) these coefficients are given in Table~\ref{tab:Wilson}.
\begin{table}[t]
\caption{The numerical input used in our analysis.}
\label{tab:param2}
\begin{center}
\begin{tabular}{ll}
\hline \hline
$|V_{tb}V_{ts}^*|=0.04026$ & $G_F=1.16637\times 10^{-5}\, {\rm GeV^{-2}}$\\
$\mu=m_b=4.8\, {\rm GeV}$   & $\alpha_{\rm em}=1/137.036$\\
$m_c=1.4\, {\rm GeV}$       &$m_B=5.27950\, {\rm GeV}$ \\
$\overline{m}_b(\mu)=4.14\, {\rm GeV}$     &$\tau_B=1.525\, {\rm ps}$ \\
$\overline{m}_s(\mu)=0.079\, {\rm GeV}$     &$m_{K^*}=0.89594\, {\rm GeV}$\\
\hline \hline
\end{tabular}
\end{center}
\end{table}
\begin{table}[th]
\caption{The SM Wilson coefficients at the scale $\mu=4.8$\,GeV,
to NNLO accuracy \cite{Altmannshofer:2009}.} \label{tab:Wilson}
\begin{center}
\begin{tabular}{ccccc}
\hline \hline
$\bar C_1(\mu)$ & ${\bar C}_2(\mu)$ &${\bar C}_3(\mu)$ &${\bar C}_4(\mu)$ &${\bar C}_5(\mu)$ \\
$-0.128$ & $1.052$ & $0.011$ & $-0.032$ &$0.009$  \\
${\bar C}_6(\mu)$ & $C_{7\gamma}^{\rm eff}(\mu)$ & $C_{8 g}^{\rm eff}(\mu)$ &$C_{9{\rm V}}(\mu)$&$C_{10{\rm A}}(\mu)$\\
$-0.037$ & $-0.304$ &$-0.167$& $4.211$&$-4.103$  \\
\hline \hline
\end{tabular}
\end{center}
\end{table}
In the numerical estimations, we use the form factors from the
light-cone sum rules (LCSR) calculation \cite{Ball:2005}.

\section{ \label{sec:results}
 Results of the calculation for the ${\bar B}_d^0\to {\bar
K}^{*0}\,e^+\,e^-$ decay }

\subsection{\label{subsec:observables}
Dependence of observables on dilepton invariant mass squared}

\begin{figure*}
\begin{center}
\includegraphics[width=.40\textwidth]{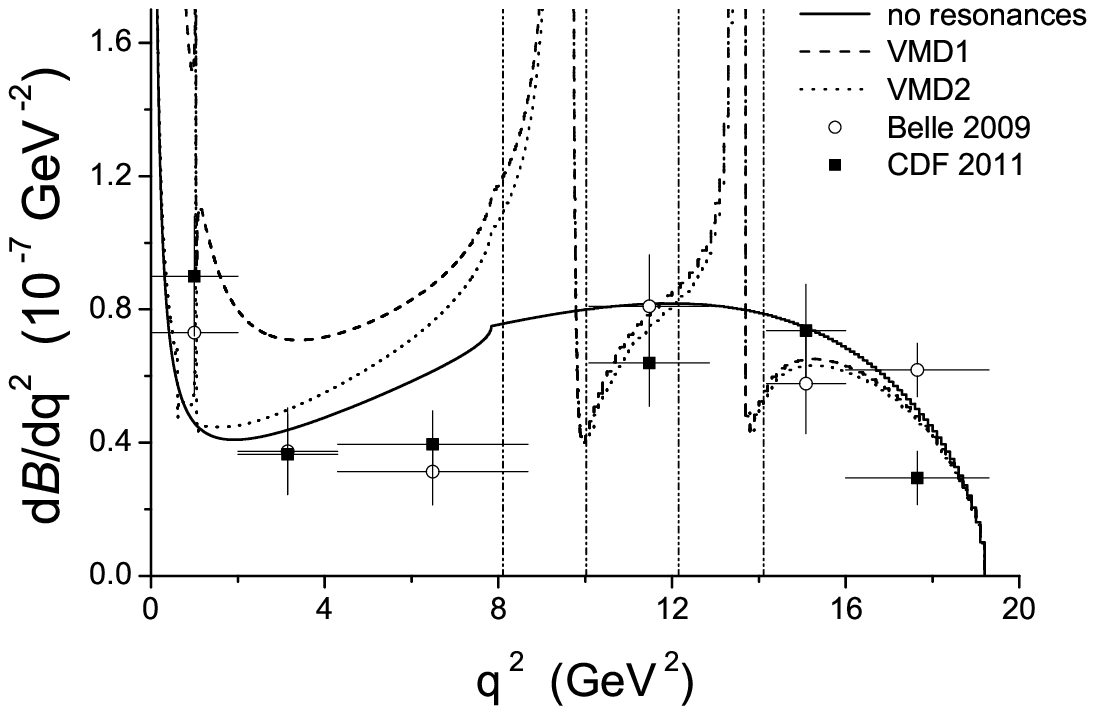}
\includegraphics[width=.40\textwidth]{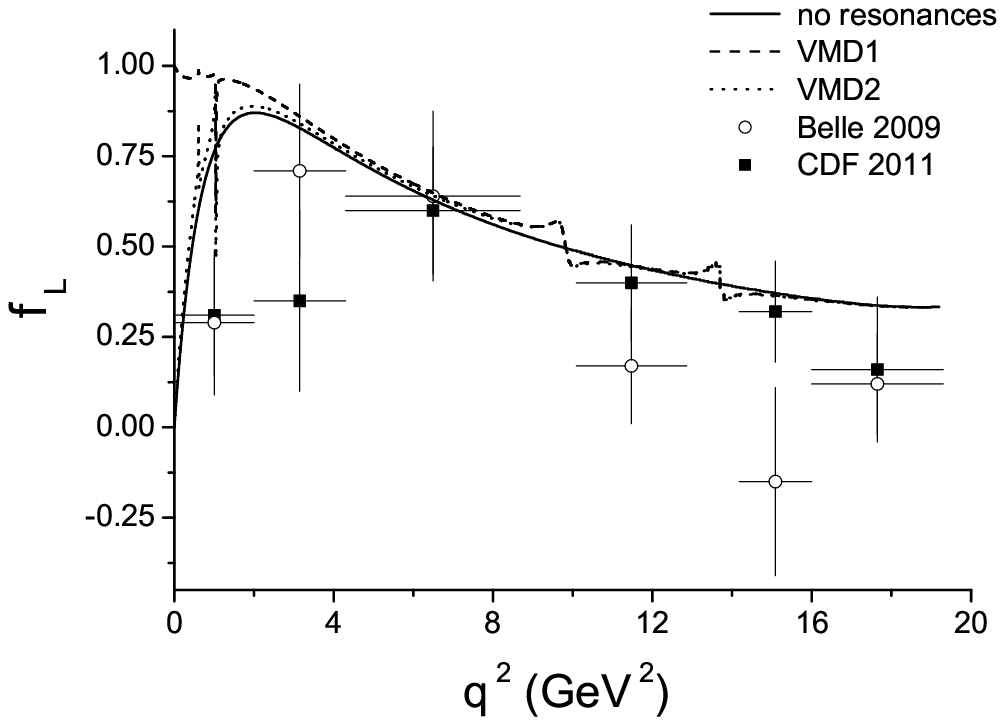}
\includegraphics[width=.40\textwidth]{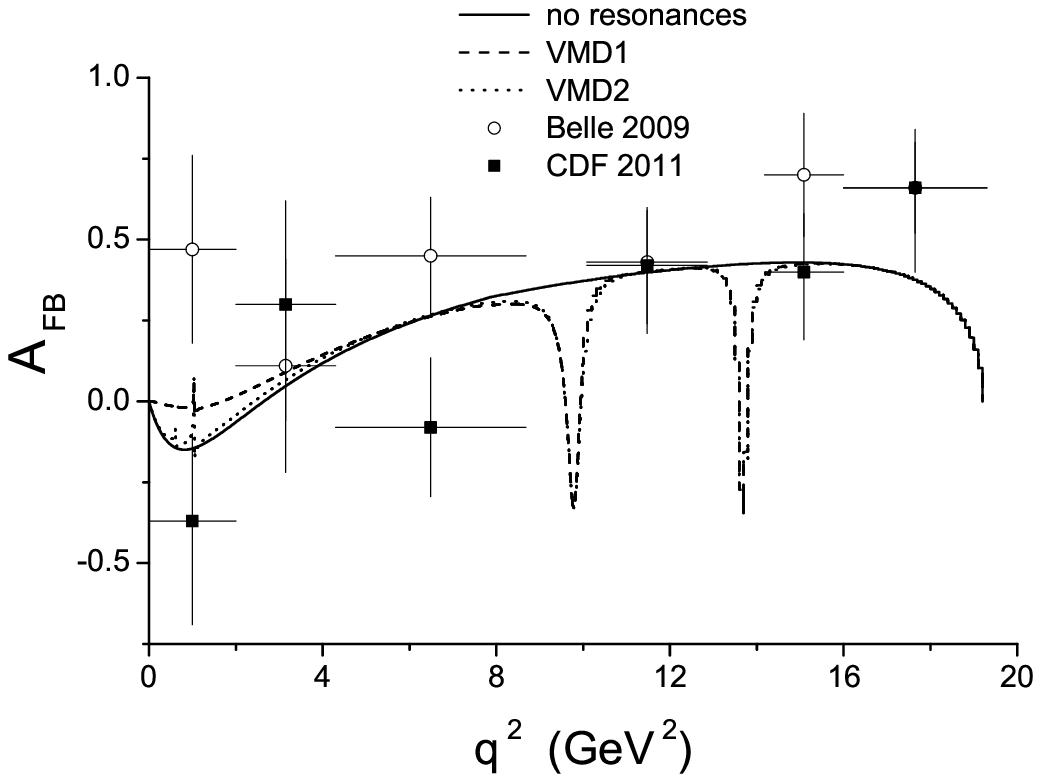}
\includegraphics[width=.40\textwidth]{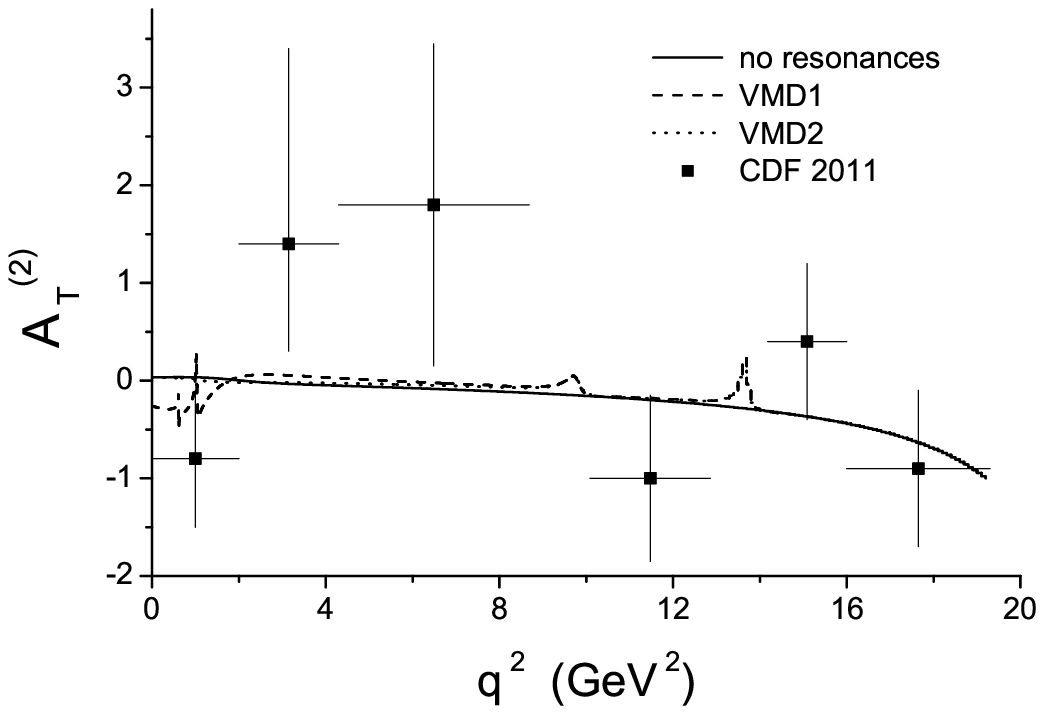}
\caption{The upper row: differential branching ratio (left), longitudinal
polarization fraction of $K^*$ meson (right); the lower row: forward-backward
asymmetry $A_{\rm FB}$ (left), coefficient $A_{\rm T}^{(2)}$ (right) as
functions of $q^2$. Solid line corresponds to the SM calculation without
resonances taken into account. Dashed and dotted lines are calculated with
account of resonances in the VMD1 and VMD2 versions of VMD model respectively.
The form factors are taken from~\cite{Ball:2005}. The data from Belle
(KEKB)~\cite{Belle:2009} and CDF (Tevatron)~\cite{CDF:2011} are shown by the
circles and boxes respectively. Due to the choice of reference frame in
Fig.~\ref{fig1}, the forward-backward asymmetry $A_{\rm FB}$ in
Refs.~~\cite{Belle:2009, CDF:2011} is related to asymmetry in
Eq.~(\ref{eq:012}) via $A_{\rm FB} = - d{\bar A}_{\rm FB}^{(l)}/d\hat{q}^2$.
The dash-double-dot vertical lines in the figure for differential branching
ratio indicate ``charmonia veto'' (see the text). } \label{fig:Br}
\end{center}
\end{figure*}

In Figs.~\ref{fig:Br} we present results for the dependence of various
observables in the ${\bar B}_d^0\to {\bar K}^{*0}\,e^+\,e^-$ decay on the
dilepton invariant mass squared. The interval of $m_{ee}=\sqrt{q^2}$ is taken
from 10 MeV to $m_B- m_{K^*} \approx 4.384$ GeV. The phase $\delta_0^{V}$ is
chosen zero for all resonances except the $\phi$ meson, for which
$\delta_0^{\phi} =2.82\, {\rm rad}$ (see Table~\ref{tab:ampl}).

As is seen in Fig.~\ref{fig:Br}, predictions of VMD1 and VMD2
models differ for $f_L$,  $A_{\rm FB}$ and $A_{\rm T}^{(2)}$ at
small $q^2 \lesssim 2$ GeV$^2$, while for the differential
branching at the bigger values, $q^2 \lesssim 8$ GeV$^2$. At the
bigger values of invariant mass, VMD1 and VMD2 yield close
results. Note that the difference between predictions of these two
models is especially large for the high-lying resonances $J/ \psi$
and $\psi(2S)$.

In addition, VMD1 and VMD2 models lead to a qualitatively different behavior of
longitudinal fraction $f_L$ at small $q^2$ (the upper right panel in
Fig.~\ref{fig:Br}). The data demonstrate that $f_L \to 0$ at very small $q^2$,
that is in agreement with calculation in the VMD2 model. In general, from
comparison with data the VMD2 version seems somewhat more preferable.

Let us comment on the coefficient $A_{\rm T}^{(2)}$ in the azimuthal
distribution (\ref{eq:013}), plotted in Fig.~\ref{fig:Br} (the lower right
panel). According to the definition (\ref{eq:014}) and due to the properties of
the $K^*$ polarization fractions \ $ 0 \leqslant f_{\|, \, \perp } \leqslant
1$, the coefficient $A_{\rm T}^{(2)}$ is constrained:
\begin{equation} -1 \leqslant A_{\rm T}^{(2)}
\leqslant 1 \,. \label{eq:041_c}
\end{equation}

The calculation in the SM with resonances yields the values of
$A_{\rm T}^{(2)}$ which are much smaller than the data (see
Fig.~\ref{fig:Br}). In this connection we note that the
experimental uncertainties are still big, and it is not clear if
the measured $q^2$-dependence of $A_{\rm T}^{(2)}$ indeed lies in
the limits (\ref{eq:041_c}).

\begin{table}[tbh]
\caption{Branching ratio for the decay ${\bar B}_d^0\to {\bar
K}^{*0}\,e^+\,e^-$ calculated within the limits: $m_{ee}^{min}
=30$ MeV, \ $m_{ee}^{max} = m_B-m_{K^*}$. The 2nd column: for the
whole interval of invariant mass, the 3rd column: with the
``charmonia veto'' (see the text).}
\begin{center}
\begin{tabular}{l c  c }
\hline \hline & \multicolumn{2}{c}{BR  ($10^{-6})$} \\
\cline{2-3}
model  & no veto &  with veto \\
\hline
SM, no res.  & 1.32   & 1.01 \\
SM, res. VMD1 & 134.2   & 49.6 \\
SM, res. VMD2 & 85.6  & 1.03 \\
\hline \hline
\end{tabular}
\end{center}
\label{tab:Branching}
\end{table}

It is seen from Fig.~\ref{fig:Br} (the upper left panel) that the
charmonia resonances contribute to the differential branching far
beyond their pole positions, especially in the VMD1 model. In
order to investigate the role of the charmonia resonances we
calculate the total branching ratio in Table~\ref{tab:Branching}.
Calculation over the whole allowed interval of invariant masses,
shown in the 2nd column, demonstrates a very big resonance
contribution.

Usually in experimental analyses~\cite{Babar:2009,Belle:2009,CDF:2011} certain
cuts are applied in order to cut out the charmonia contributions (the so-called
charmonia veto). In the 3rd column of Table ~\ref{tab:Branching} we used the
integration region with the following cut out intervals taken as in the BaBar
analysis~\cite{Babar:2009} for the $e^+ e^-$ pairs: \ $ 8.11 \leqslant q^2
\leqslant 10.03$ GeV$^2$ and $ 12.15 \leqslant q^2 \leqslant 14.11$ GeV$^2$.

As is seen, the $c \bar{c}$-resonances contribution is completely
eliminated. The result obtained in the VMD2 model becomes close
(within a few percent) to the calculation without resonances. At
the same time the VMD1 calculation still yields a big value of
branching ratio, which is due to the steep rise of the
differential branching at small invariant masses in
Fig.~\ref{fig:Br}.

The calculations, presented in Table~\ref{tab:Branching}, are the
predictions of our model for the current experiments carried out
at LHCb~\cite{Lefrancois:2009}. We can also compare predictions
(with veto) in Table~\ref{tab:Branching} with experimental
measurements: \ $(1.07^{+0.11}_{-0.10} \pm{0.09}) \times 10^{-6}$
for the $B \to K^{*} \, \ell^+\,\ell^-$ decay ($\ell = e, \, \mu$)
at Belle~\cite{Belle:2009}, $(1.02 \pm 0.10 \pm 0.06) \times
10^{-6}$ for the $ B^0 \to K^{*0}\, \mu^+\,\mu^-$ decay at
CDF~\cite{CDF:2011}, and $(1.02^{+0.30}_{-0.28} \pm{0.06}) \times
10^{-6}$ for the $B^0 \to K^{*0}\, e^+\, e^-$ decay at
BaBar~\cite{Babar:2009}.
\begin{figure*}
\begin{center}
\includegraphics[width=.32\textwidth]{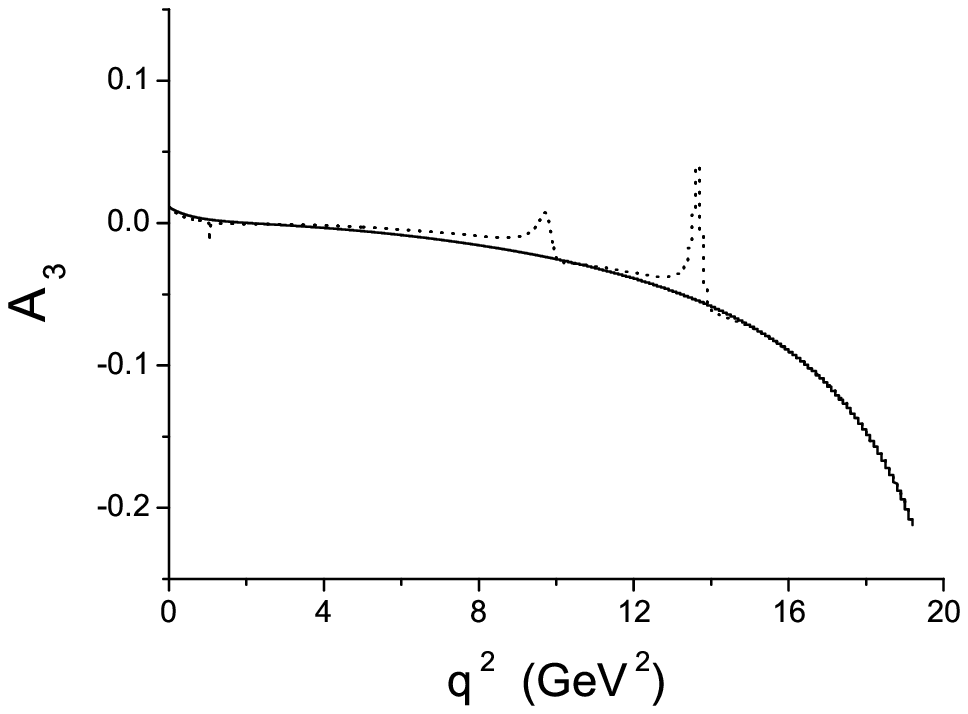}
\includegraphics[width=.32\textwidth]{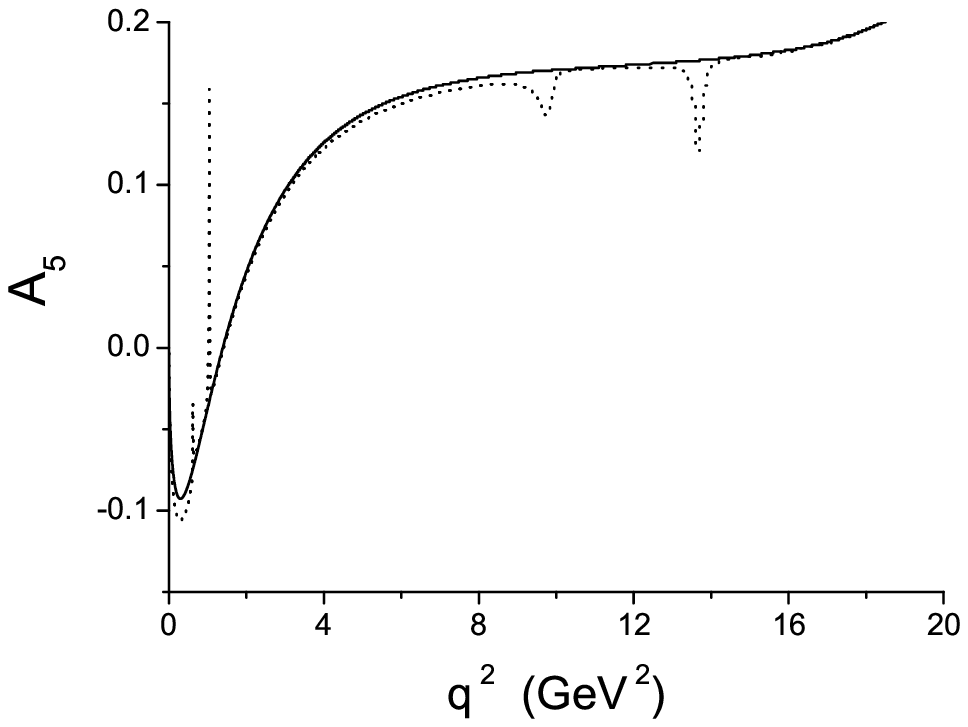}
\includegraphics[width=.32\textwidth]{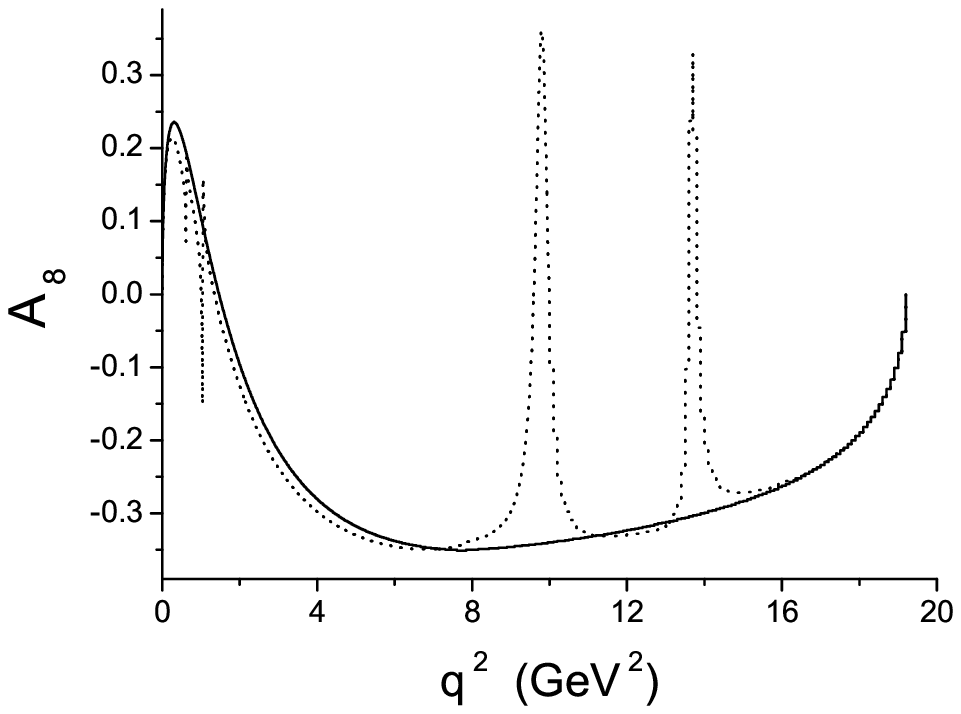}
\includegraphics[width=.32\textwidth]{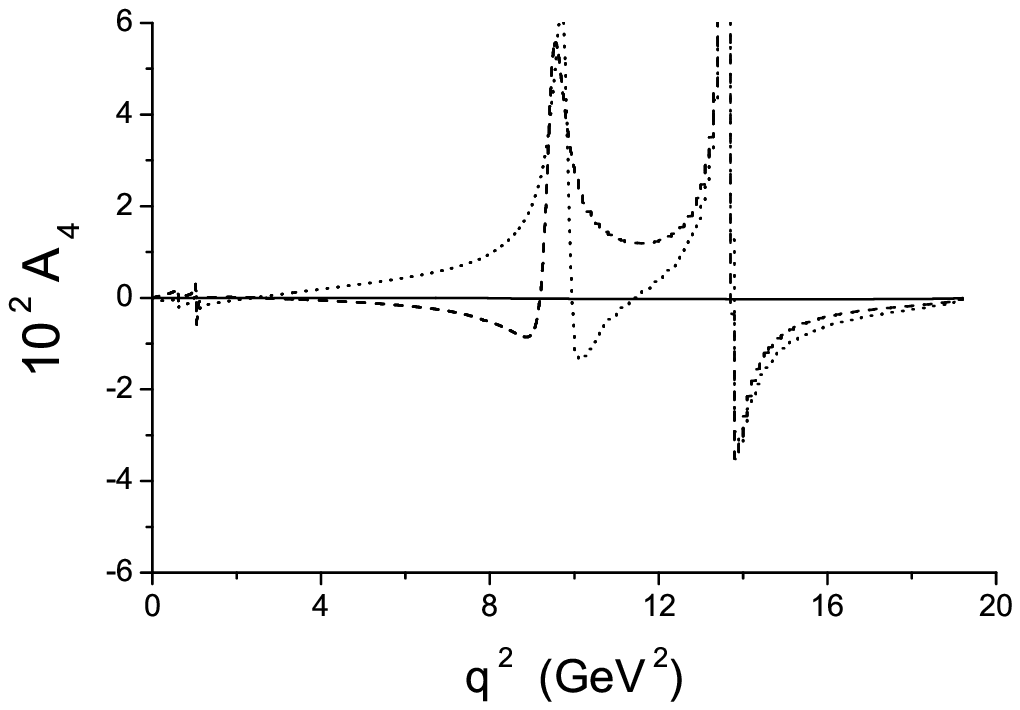}
\includegraphics[width=.32\textwidth]{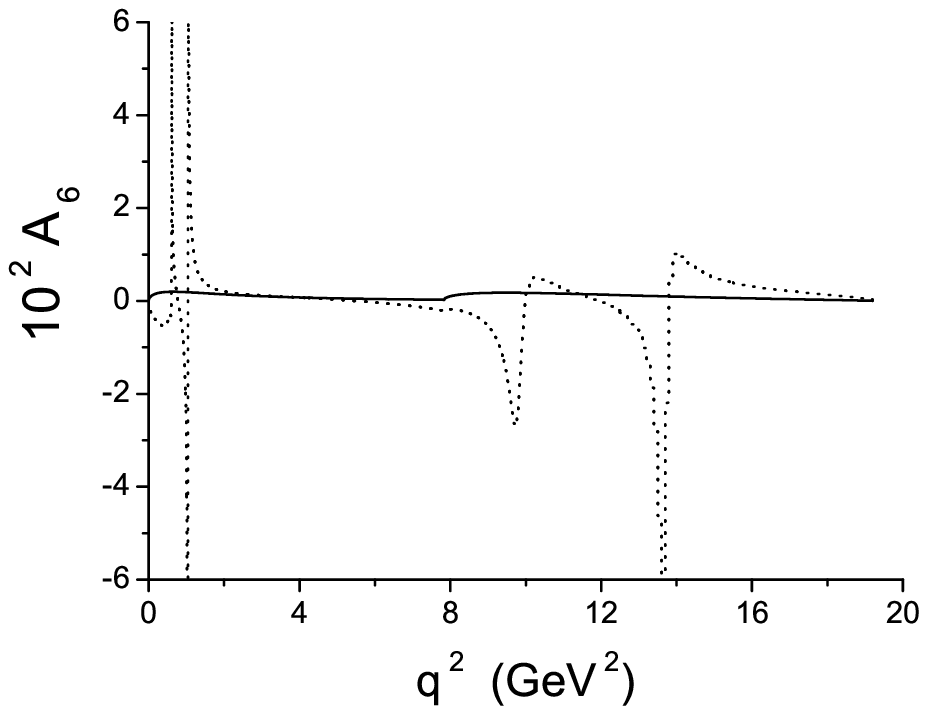}
\includegraphics[width=.32\textwidth]{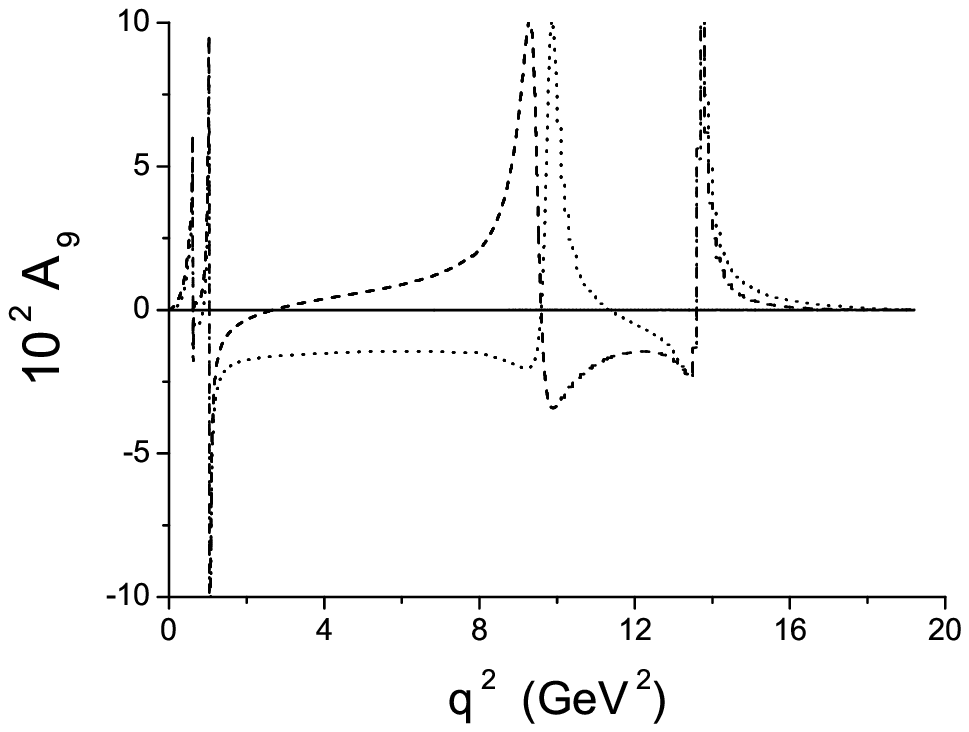}
\caption{Asymmetries as functions of $q^2$. The upper row from
left to right: $A_3$, $A_5$ and $A_8$; the lower row from left ro
right: $A_4$, $A_6$ and $A_9$. Solid lines correspond to the SM
calculation without resonances, dotted (dashed) lines correspond
to calculation including resonances with the zero-helicity phase
$\delta_0^{J/\psi}=0$ ($\delta_0^{J/\psi}= 2.45$ rad). Calculation
is performed in the VMD2 version. } \label{fig:A3}
\end{center}
\end{figure*}

In Figs.~\ref{fig:A3} we present asymmetries calculated according
to Eqs.~(\ref{eq:014})--(\ref{eq:016}), (\ref{eq:018}),
(\ref{eq:019}), (\ref{eq:021}), (\ref{eq:022}).

Asymmetry $A_3 = \frac{1}{\pi}(1-f_L) A_{\rm T}^{(2)}$ takes
sizable values at large invariant masses, while in the wide region
of $m_{ee}$ this observable is small, of the order of $10^{-2}$
(see also Figs.~\ref{fig:Br} for $A_{\rm T}^{(2)}$). Account of
resonances changes it mainly in the vicinity of the resonance
positions, i.e. at $m_{ee} \approx m_V$.

As for $A_5$ and $A_8$, they take sizable values in the whole
region of $q^2$ (see Figs.~\ref{fig:A3}). The resonances give
considerable contribution, especially to $A_8$. One of features is
the point $q^2_0$, where these asymmetries cross zero. Calculation
shows that the zero point, $q^2_0 \sim 1.5$ GeV$^2$, is almost
insensitive to the presence of the resonances. This feature makes
these asymmetries convenient observables for experimental study,
similarly to the forward-backward asymmetry $A_{{\rm FB}}$ in
Fig.~\ref{fig:Br}.

Some of the asymmetries are very small in the SM without
resonances, in particular, $A_9  \equiv 0$, $A_4 \sim 10^{-4}$ and
$A_6 \sim 10^{-3}$ (Figs.~\ref{fig:A3}). Note that these
asymmetries are determined by the imaginary part of the bilinear
combinations of the amplitudes. The imaginary part of the
amplitudes in the SM (without resonances) is determined by the
$u,\, d,\, s$ and $c$ quark loop through the function $Y(q^2)$
\cite{BFS:2001}, and therefore the imaginary part of the
nonresonant amplitudes appears to be very small. Therefore in
framework of the naive factorization, applied in the present work,
it is not surprising that these asymmetries are small in the SM
without resonances. Inclusion of the resonances changes behavior
of these asymmetries in the wide region of invariant masses (see
Figs.~\ref{fig:A3}).

Recently the asymmetry $A_{{\rm Im}}= -\frac{\pi}{2}A_4$ in the
$B^0 \to K^{*0}\, \mu^+\,\mu^-$ decay has been measured for the
first time at CDF~\cite{CDF:2011}. Note that the average values of
this asymmetry over the $q^2$-ranges $[0.0 - 4.3)$ GeV$^2$ and
$[1.0 - 6.0)$  GeV$^2$ are consistent with our calculations in
framework of the SM.

One should note, that the amplitudes for the decay of $B$ meson to
two vector mesons are experimentally determined by the four
polarization parameters, branching fraction and one overall phase
$\delta_0^V $ (the phase of the amplitude with zero helicity for
decay $B \to K^* V$). As it is seen from
Eqs.~(\ref{eq:038})--(\ref{eq:040}) the contribution of resonances
depends on the invariant amplitudes $S^{V}_i$. Values of these
amplitudes are determined in Table~\ref{tab:ampl}, however their
phases are defined with respect to the phase $\delta_0^V$ , which
is experimentally known only for the decay $B \to K^* \phi$. For
other resonances, the phase $\delta_0^V$ is not known either
experimentally or theoretically.

As is seen in Figs.~\ref{fig:A3}, the asymmetries $A_4$ and $A_9$ essentially
depend on the choice of the $\delta_0^{J/\psi}$ phase. Thus, in order to
unambiguously determine the resonance contribution to the process ${\bar B}_d^0
\to {\bar K}^{*0} \, (\to K^{-}\, \pi^+) \, \ell^+\, \ell^- $, the phases
$\delta_0^V$ should be known for all vector resonances $\rho, \omega, \phi,
J/\psi, \ldots $. These phases can be found from experiments on $B$-meson
decays through the interference with other $B$ decays with the same final
states, for example, $B^0 \to \phi K^{*0}$ and $B^0 \to \phi (K \pi)_0^{*0}$,
where $(K \pi)_0^{*0}$ is the $J^P =0^+$  \ $K \pi$
component~\cite{Babar:2007}.


\subsection{\label{subsec:two_approaches}
Comparison of two approaches to including the resonances }

Earlier in the literature \cite{Deshpande:1989} it has been
suggested to combine the factorization assumption and VMD
approximation in estimating LD effects for the $B$ decays. This
can be accomplished in an approximate manner through the
substitution
\begin{equation}\label{eq:042}
C_{9V}^{\rm eff}\to C_{9V}^{\rm
eff}-\frac{3\,\pi}{\alpha_{em}^2}C^{(0)}\sum_{\psi} k_{\psi}
\frac{\hat{m}_{\psi}\,\hat{\Gamma}(\psi\to e^+e^-)}{\hat{q}^2 -
\hat{m}_{\psi}^2 + i\hat{m}_{\psi} \hat{\Gamma}_{\psi}}\,,
\end{equation}
where the properties of the vector mesons
$\psi=J/\psi(1S)\,,\psi(2S)\,,\ldots,\psi(4415)$ are summarized in
Table~\ref{tab:param1}, \ $\hat{\Gamma}_{\psi\to e^+e^-}\equiv
\Gamma_{\psi\to e^+e^-}/m_B$, \ $C^{(0)}=3\,\bar C_1+\bar
C_2+3\,\bar C_3+\bar C_4+3\,\bar C_5+\bar C_6$, and the Wilson
coefficients are presented in Table~\ref{tab:Wilson}.

The last term in Eq.~(\ref{eq:042}) describes the LD contribution
of the real intermediate $\bar c\,c$ states,
$k_\psi=|k_\psi|\,e^{i\delta_\psi}$ is the factor that the $B\to
K^* \psi$ amplitude, calculated using naive factorization, must be
multiplied by to get the measured $B\to K^* \psi$ rate. Under
naive factorization, the branching ratio for $B \to K^* \psi$ is
\begin{eqnarray}\label{eq:043}
{\rm BR}(B \to K^* \psi) &=&m_B\,\tau_B\,
\frac{\sqrt{\lambda(1,\hat{m}_{K^*}^2,\hat{m}_\psi^2)}}{16\pi}\nonumber \\
& \times & \left(\frac{G_F m_B^2}{\sqrt{2}}\right)^2\,
\left|V_{tb}V_{ts}^*\frac{C^{(0)}}{3}\right|^2 \nonumber \\
& \times &\left( |X_0^\psi|^2 + |X_\|^\psi|^2 + |X_\perp^\psi|^2
\right) \,,
\end{eqnarray}
where
\begin{eqnarray}\label{eq:044}
&& X_0^\psi=\frac{\hat{f_\psi}}{2\,\hat{m}_{K^*}(1+\hat{m}_{K^*})}
\Bigl((1+\hat{m}_{K^*})^2
(1-\hat{m}_{K^*}^2 \nonumber \\
&&-\hat{m}_\psi^2)\,A_1(m_\psi^2) -
\lambda(1,\hat{m}_{K^*}^2,\hat{m}_\psi^2)\,A_2(m_\psi^2) \Bigr)
\,,
\end{eqnarray}
\begin{equation}\label{eq:045}
X_\|^\psi=-\sqrt{2}\,\hat{m}_\psi\hat{f_\psi}(1+\hat{m}_{K^*})\,A_1(m_\psi^2)
\,,
\end{equation}
\begin{equation}\label{eq:046}
X_\perp^\psi=\sqrt{2\,\lambda(1,\hat{m}_{K^*}^2,\hat{m}_\psi^2)}\,\hat{m}_\psi\hat{f_\psi}/(1+\hat{m}_{K^*})\,V(m_\psi^2)
\,.
\end{equation}
Here, $\hat{f_\psi}\equiv f_\psi/m_B$. \

We calculate the values of $k_\psi$ for each $c \bar c$ resonance
using experimental information from Tables~\ref{tab:param1} and
\ref{tab:ampl} and equation
\begin{equation}
|k_\psi|^2 = \frac{{\rm BR}(B \to K^* \psi)_{exp} }{{\rm BR}(B \to
K^* \psi)_{theor}} \,. \label{eq:047}
\end{equation}
Using the form factors $A_1, \, A_2, \, V$ in the LCSR
model~\cite{Ball:2005} we find the values: \ $|k_{J/\psi}|=0.894$,
\ $|k_{\psi(2S)}|=0.841$ and for the higher resonances the average
of the $|k_{J/\psi}|$ and $|k_{\psi(2S)}|$ is used. The phase of
$k_\psi$ is chosen zero as in the factorization approach. Note,
that the above values of the parameters $k_\psi$ are considerably
smaller that the values used in the earlier
papers~\cite{Deshpande:1989,Ligeti:1996,Ali:2000}. Therefore we
can expect the smaller resonance contribution to the differential
branching and forward-backward asymmetry as compared with results
of these papers.

Using Eqs.~(\ref{eq:042}), (\ref{eq:023})--(\ref{eq:025}) we
calculate the observables for the ${\bar B}_d^0 \to {\bar K}^{*0}
e^+ e^-$ decay. These results are compared with results of
calculations performed in the previous sections, in which only
$J/\psi$ and $\psi(2S)$ resonances are included. This comparison
may show sensitivity of the branching, $K^*$ polarization
fractions and asymmetries to the method of including the vector
resonances.

\begin{figure*}
\begin{center}
\includegraphics[width=.32\textwidth]{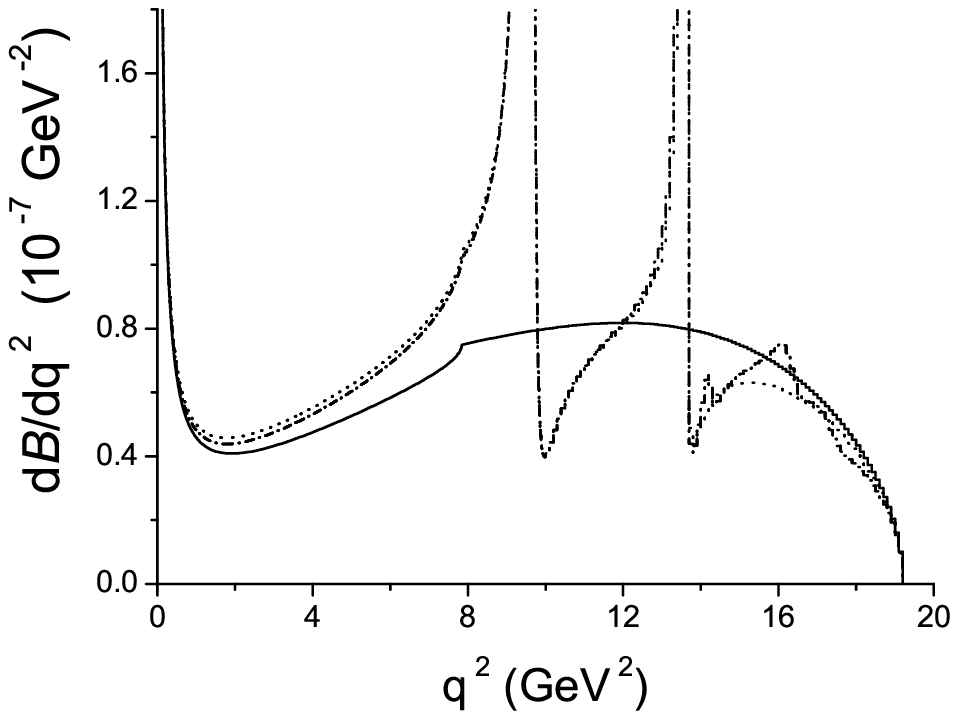}
\includegraphics[width=.32\textwidth]{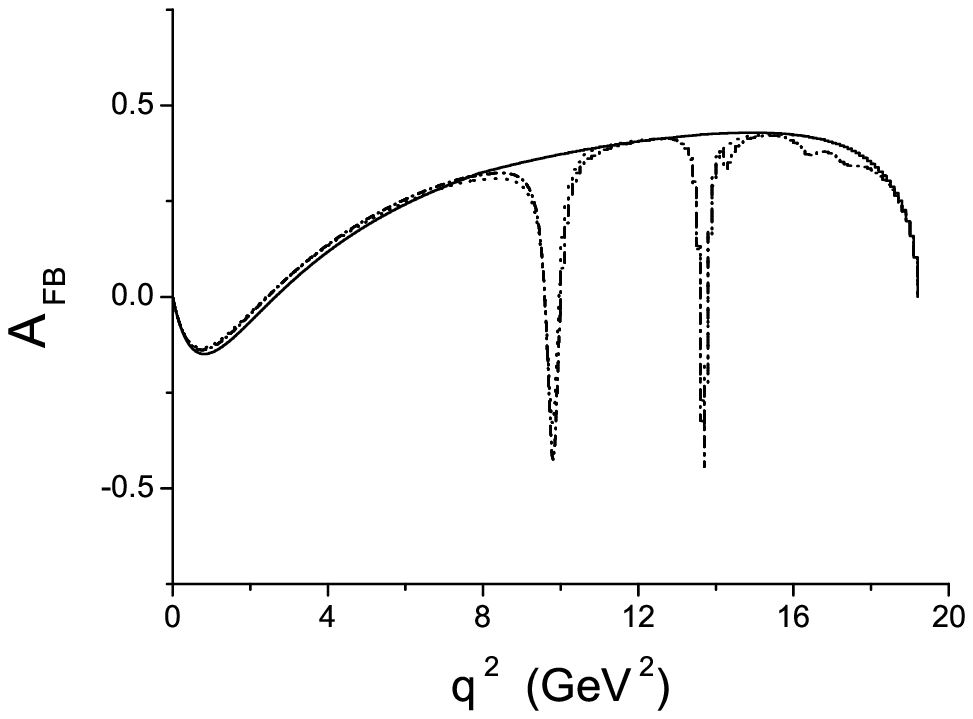}
\includegraphics[width=.32\textwidth]{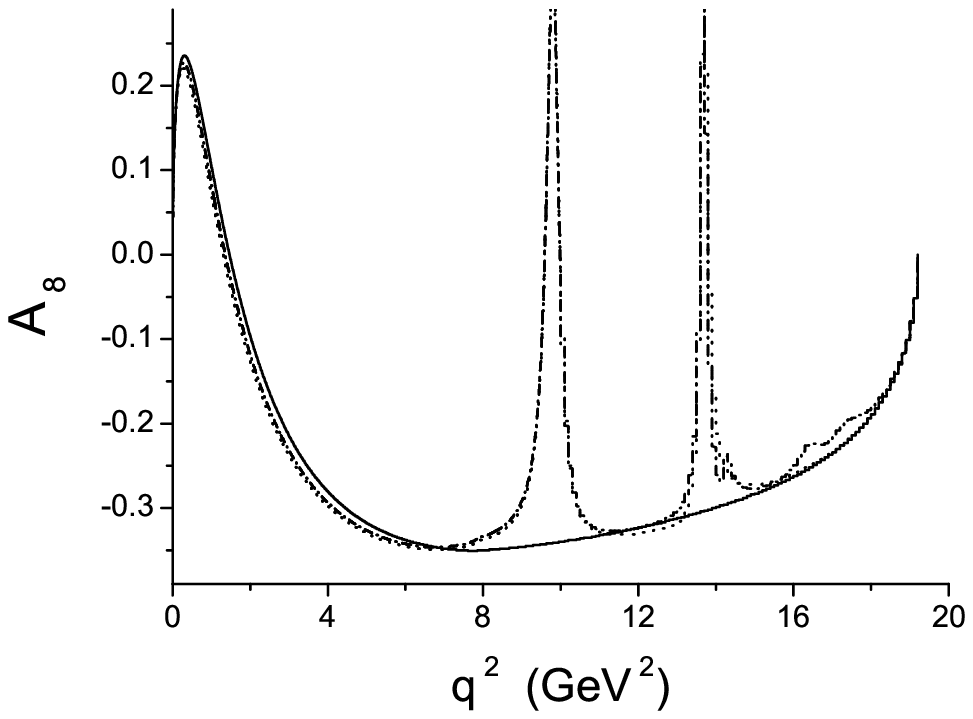}
\end{center}
\caption{Differential branching (left), forward-backward asymmetry
(middle) and asymmetry $A_8$ (right) a function of $q^2$. Solid
lines are calculated without resonances, dotted lines are
calculated with resonances $J/\psi$, $\psi(2S)$ according
Eqs.~(\ref{eq:038})--(\ref{eq:040}), and dash-dotted lines with
resonances according to Eqs.~(\ref{eq:042}),
(\ref{eq:023})--(\ref{eq:025}). } \label{fig:01res}
\end{figure*}

\begin{figure*}
\begin{center}
\includegraphics[width=.32\textwidth]{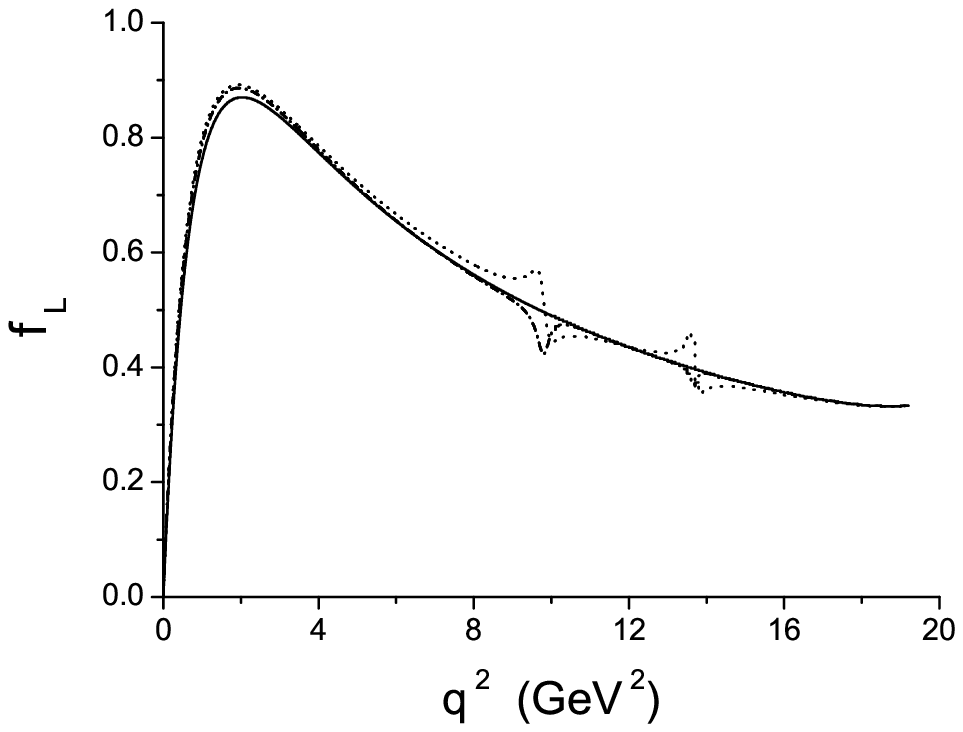}
\includegraphics[width=.32\textwidth]{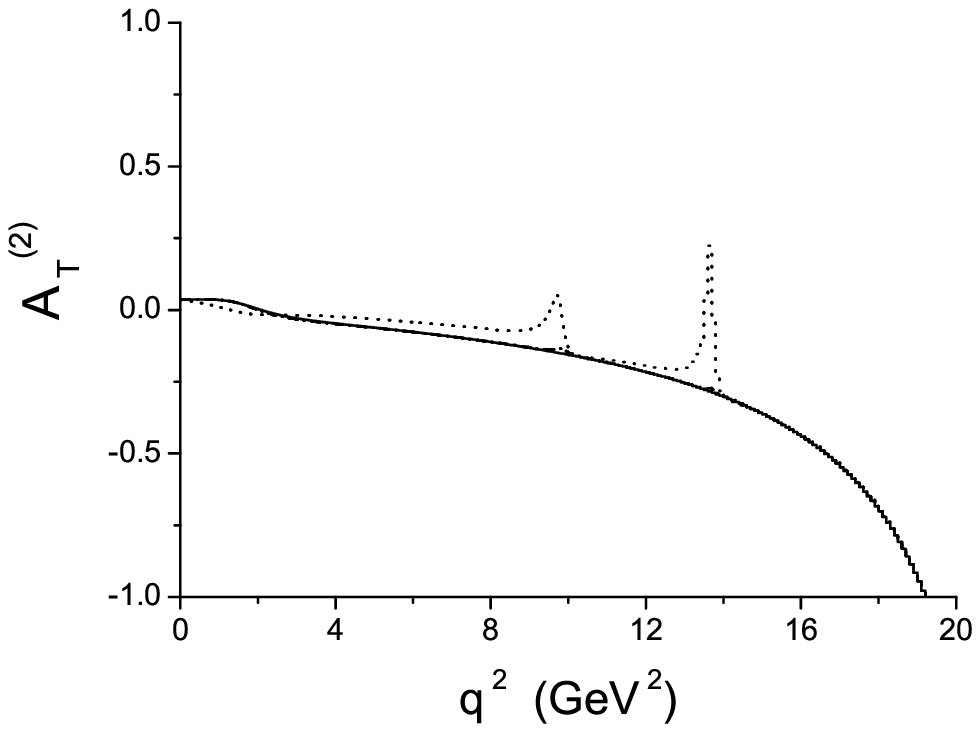}
\includegraphics[width=.32\textwidth]{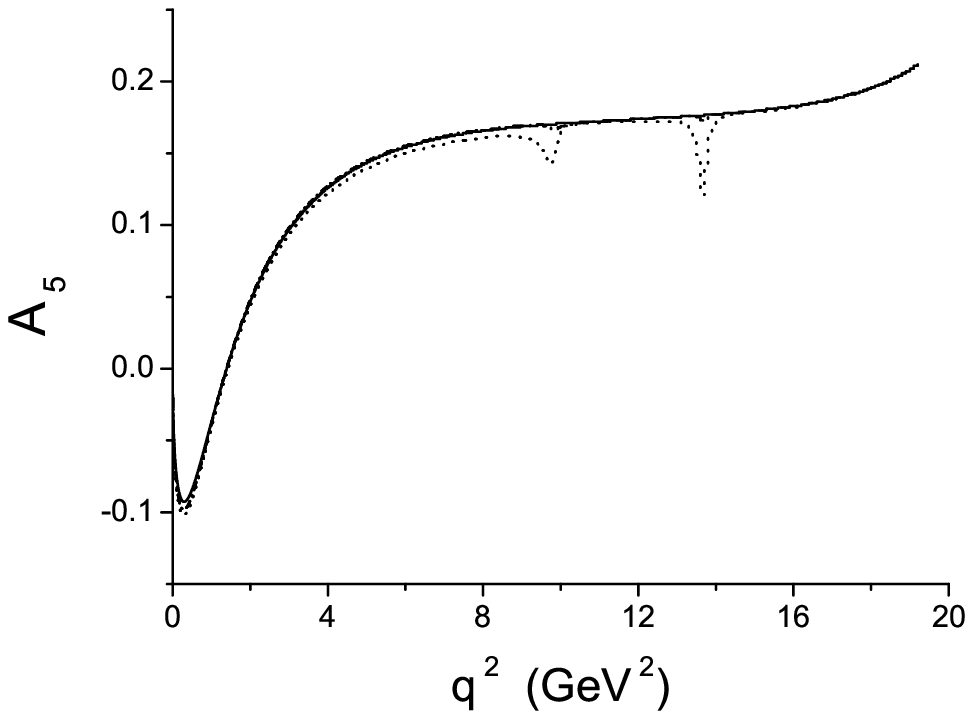}
\includegraphics[width=.32\textwidth]{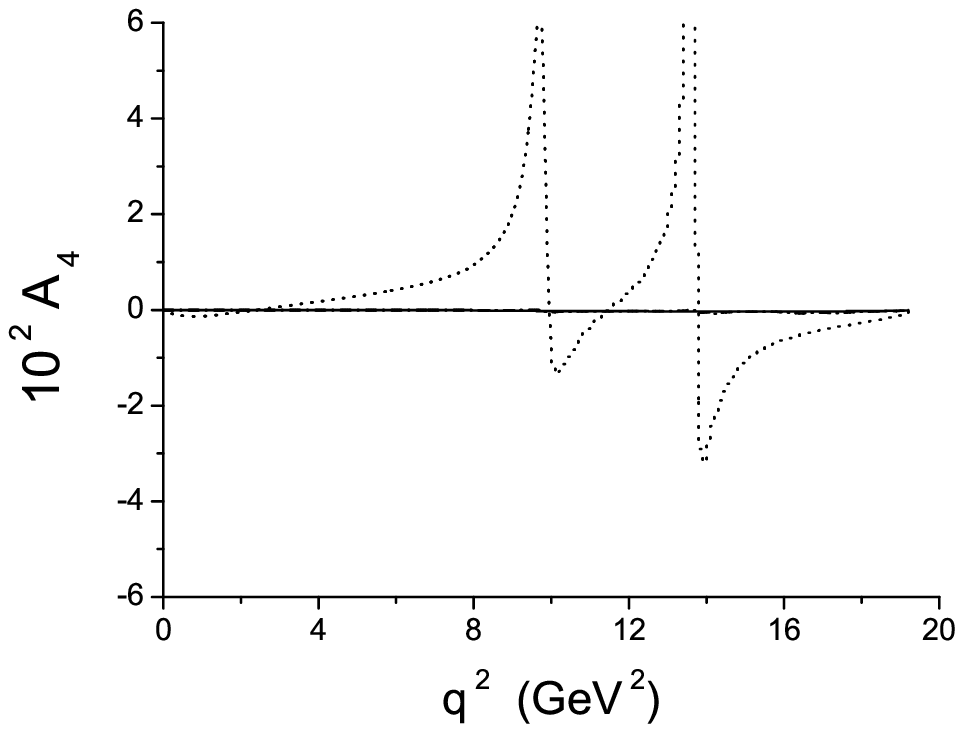}
\includegraphics[width=.32\textwidth]{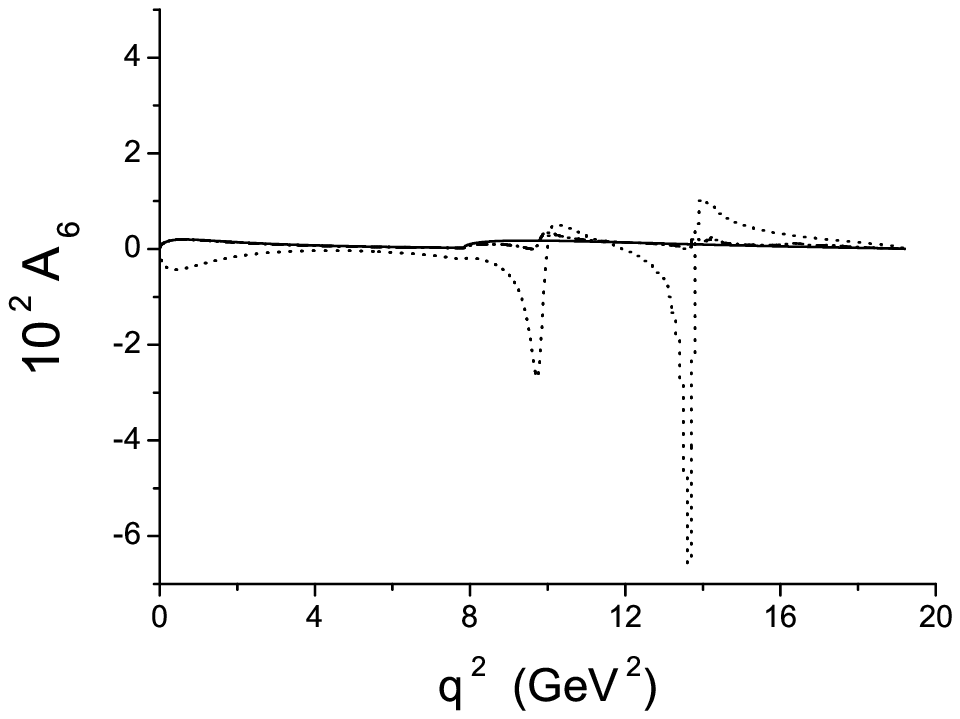}
\includegraphics[width=.32\textwidth]{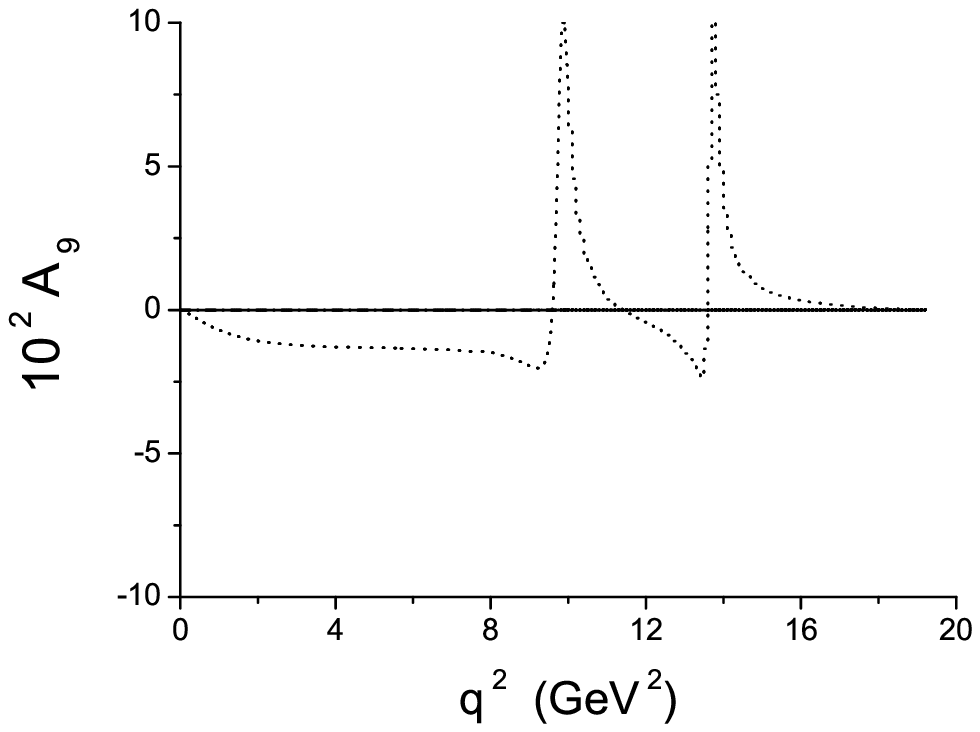}
\end{center}
\caption{The upper row from left to right: longitudinal
polarization fraction $f_L$, asymmetries $A_{\rm T}^{(2)}$ and
$A_5$;  the lower row from left ro right: asymmetries $A_4$, $A_6$
and $A_9$, as functions of $q^2$. Solid lines are calculated
without resonances, dotted lines are calculated with resonances
$J/\psi$, $\psi(2S)$ according
Eqs.~(\ref{eq:038})--(\ref{eq:040}), and dash-dotted lines with
resonances according to Eqs.~(\ref{eq:042}),
(\ref{eq:023})--(\ref{eq:025}). } \label{fig:02res}
\end{figure*}

Firstly, from Fig.~\ref{fig:01res} one sees that the two ways of
including the resonances give very close predictions for
differential branching, forward-backward asymmetry and asymmetry
$A_8$ in the region of $q^2$ up to $m^2_{\psi(2S)}$. Note, that
calculations of the branching and forward-backward asymmetry have
been performed in Ref.~\cite{Ali:2000} using formulas analogous to
Eqs.~(\ref{eq:042}). Our results in Figs.~\ref{fig:01res}
qualitatively agree with calculations in \cite{Ali:2000}.

Secondly, in Fig.~\ref{fig:02res} we present asymmetries, which strongly depend
on the method of including resonances. As for asymmetries $A_{\rm T}^{(2)}$ and
$A_5$, these two methods give essentially different predictions in the region
of resonances $J/\psi$ and $\psi(2S)$, while far from the resonance region, the
predictions coincide with the non-resonant calculations. In the other
asymmetries, $A_4$, $A_6$ and $A_9$, the two methods give different predictions
not only in the resonance region but also at $q^2$ away from resonances.

\section{ \label{sec:conclusions} Conclusions}

The rare FCNC decay ${\bar B}_d^0 \to {\bar K}^{*0} \, (\to K^{-}\, \pi^+) \,
e^+\,e^- $ has been studied in the whole region of electron-positron invariant
masses up to $m_B -m_{K^*}$. The fully differential angular distribution over
the three angles and dilepton invariant mass  for the four-body decay ${\bar
B}_d^0 \to K^{-}\, \pi^+ \, e^+\,e^- $ is analyzed. We defined a convenient set
of asymmetries which allows one to extract them from measurement of the angular
distribution once sufficient statistics is accumulated. We performed
calculations of the differential branching ratio, polarization fractions of
$K^*$ meson and asymmetries. These asymmetries may have sensitivity to various
effects of NP, although in order to see signatures of these effects the
resonance contribution should be accurately estimated.

Contribution of the intermediate vector resonances in the process $\bar{B}_d^0
\to \bar{K}^{*0} \, (\to K^{-}\, \pi^+) \, V $ with $V = \rho(770), \,
\omega(782), \, \phi(1020), \, J/\psi, \, \psi(2S)$, decaying into the $e^+
e^-$ pair, has been taken into account. Various aspects of theoretical
treatment of this long-distance contribution have been studied.

The important aspect is the choice of the VMD model, describing
the $V \, \gamma$ transition. We used two variants of the VMD
model, called here VMD1 and VMD2 versions, in particular, the VMD2
one is explicitly gauge-invariant and yields a $V \, \gamma$
vertex which is suppressed at small values of the photon invariant
mass far from the vector-meson mass shell. This is especially
important in the treatment of the high-lying $J/\psi$ and other $c
\bar c$ resonances at values $q^2 \ll m_V^2$. Some of the
observables appeared to be rather sensitive to the choice of the
$V \, \gamma$ vertex. Based on comparison of calculation with the
recent data from Belle and CDF experiments we can conclude that
the VMD2 version is somewhat more preferable.

For the vertex $\bar{B}_d^0 \to \bar{K}^{*0} \, V $ we used an off-mass-shell
extension of the helicity amplitudes describing production of on-shell vector
mesons. For the latter the experimental information is used if available, and
otherwise theoretical predictions.

The total branching ratio of the decay ${\bar B}_d^0 \to {\bar
K}^{*0} \, e^+\,e^- $ in the interval $30 \, {\rm MeV} \leq m_{ee}
\leq m_B-m_{K^*}$ is calculated, and the resonance contribution is
evaluated. The latter appeared to be very big, as expected. The
calculated branching is ${\rm BR} = 1.32 \times 10^{-6} $ ($85.6
\times 10^{-6})$ without resonances (with resonances in VMD2
version). To cut out the $c \bar{c}$-resonance contribution, we
also applied the ``charmonia veto'' as is usually done in
experimental analyses~\cite{Babar:2009,Belle:2009,CDF:2011}. Then
our prediction for the total branching becomes ${\rm BR} = 1.01
\times 10^{-6} $ ($1.03 \times 10^{-6})$ for the SM calculation
without resonances (with resonances in VMD2 model).

All asymmetries are calculated in the whole region of invariant
masses. The polarization asymmetry $A_{\rm T}^{(2)}$ (and $A_3 =
\frac{1}{\pi}(1-f_L) A_{\rm T}^{(2)}$) takes sizable values only
at large $m_{ee}$. Account of resonances changes $A_{\rm T}^{(2)}$
mainly in the vicinity of the resonances, i.e. at $m_{ee} \approx
m_V$.
The asymmetries $A_5$ and $A_8$ take big values in the whole region of $q^2$,
and resonances noticeably contribute, especially to $A_8$. An interesting
feature of these asymmetries is their crossing zero at some $q^2_0$. This zero
point, $q^2_0 \sim 1.5$ GeV$^2$, turns out to be almost independent of the
presence of resonances, and this property makes these asymmetries convenient
observables for experimental study, similarly to the forward-backward asymmetry
$A_{\rm FB}$.

Some of the asymmetries are very small in the SM without
resonances, in particular, $A_9  \equiv 0$, \ $A_4 \sim 10^{-4}$
and $A_6 \sim 10^{-3}$; inclusion of the resonances changes
behavior of these asymmetries considerably.

Our calculations are compared with recent
data~\cite{Belle:2009,CDF:2011} for $q^2$-dependence of the
differential branching ratio, longitudinal polarization fraction
of $K^*$, forward-backward asymmetry $A_{{\rm FB}}$ and transverse
asymmetry $A_{{\rm T}}^{(2)}$. On the whole, results of
calculation are in reasonable agreement with the data.

We also compared predictions of our method with other existing in
the literature method of including the $c \bar c$ resonances
through a modification of the Wilson coefficient
$C_{9V}^{eff}$~\cite{Deshpande:1989}.
Our calculation shows that a few observables (differential
branching ratio, forward-backward asymmetry and asymmetry $A_8$)
are independent of the calculation method, if the parameters
$|k_\psi|$ in Eq.~(\ref{eq:042}) are equal to: $0.894$ for
$J/\psi$, \ 0.841 for $\psi(2S)$ and 0.8675 for the higher $c \bar
c$ resonances. The phase of $k_\psi$ is chosen zero. These values
are considerably smaller that the values used
in~\cite{Deshpande:1989,Ligeti:1996,Ali:2000}.

At the same time there exist asymmetries, predictions for which
are substantially different in these two methods, namely, $A_{\rm
T}^{(2)}$, \ $A_5$ are different in the vicinity of resonances
$J/\psi$ and $\psi(2S)$, while $A_4$, \ $A_6$, \ $A_9$ are
different in the whole region of invariant masses. Thus
measurement of the latter asymmetries may also be useful for
selecting a more adequate method of including the long-distance
resonance contribution to the ${\bar B}_d^0 \to {\bar K}^{*0} \,
(\to K^{-}\, \pi^+) \, \ell^+\,\ell^- $ decay.

Calculations performed in the present work may be useful for experiments aiming
at search of effects of the NP in the decay ${\bar B}_d^0 \to {\bar K}^{*0} \,
(\to K^{-}\, \pi^+) \, \ell^+\,\ell^- $.

\appendix

\section{\label{sec:Appendix} Matrix element and form factors}

\subsection{\label{subsec:matrix element} Matrix element}

The effective Hamiltonian for the quark-level transition $ b \to
s\, e^+ e^- $ within the SM is well-known and can be taken, e.g.,
from Ref.~\cite{Antonelli:2009}. It is expressed in terms of the
local operators ${\cal O}_i$ and Wilson coefficients $C_i$, where
$i=1, \ \ldots, \ 6, \ 7\gamma,\ 8g,\ 9V, 10A$.

The matrix element of this effective Hamiltonian for the
nonresonant decay ${\bar B}_d^0 (p)\to {\bar
K}^{*0}(k,\epsilon)\,e^+(q_+)\,e^-(q_-)$ can be written, in the
so-called naive factorization~\cite{Antonelli:2009}, as
\begin{eqnarray}
{\cal M}_{\rm NR} &=& \frac{G_F\alpha_{\rm
em}}{\sqrt{2}\pi}\,V_{ts}^* V_{tb}\Bigl(\langle{\bar
K}^{*0}(k,\epsilon)|\bar{s} \gamma_\mu P_L b|{\bar
B}_d^0(p)\rangle \nonumber \\ &&\times\bigl(C_{9V}^{\rm eff}
\bar{u}(q_-) \gamma^\mu v(q_+) +C_{10A}\bar{u}(q_-) \gamma^\mu
\gamma_5 v(q_+)\bigr)\nonumber \\ && -\frac{2}{q^2}
\overline{m}_b(\mu) \langle{\bar K}^{*0}(k,\epsilon)|\bar{s}\,i\,
\sigma_{\mu\nu} q^\nu( C_{7\gamma}^{\rm eff} P_R \nonumber \\ &&+
C_{7\gamma}^{\prime \, \rm eff} P_L)\, b|{\bar
B}_d^0(p)\rangle\,\bar{u}(q_-) \gamma^\mu
v(q_+)\Bigr)\,.\label{eq:0A1}
\end{eqnarray}

Here, $P_{L,R}=(1\mp \gamma_5)/2$ denote chiral projectors, and
$\overline{m}_b(\mu)$ [$\overline{m}_s(\mu)$] is the running
bottom (strange) quark mass in the $\overline{\rm MS}$ scheme at
the scale $\mu$. Moreover, $\sigma_{\mu \nu}={i\over
2}[\gamma_{\mu},\gamma_{\nu}]$, $q_{\mu}=(q_{+} +q_{-})_{\mu}$,
$C_{7\gamma}^{\rm eff} = C_{7\gamma}-(4 \bar C_3-\bar C_5)/9-(4
\bar C_4-\bar C_6)/3$, $C_{9V}^{\rm eff} = C_{9V} +Y(q^2)$, where
$Y(q^2)$ is given in Ref.~\cite{BFS:2001}. Note that in the
framework of the SM  \ $\overline{m}_b(\mu) \,
C_{7\gamma}^{\prime\,\rm eff}= \overline{m}_s (\mu)
\,C_{7\gamma}^{\rm eff}$.

The ``barred'' coefficients $\bar{C}_i$ (for $i=1,\ldots,6$) are
defined as certain linear combinations of the $C_i$, such that the
$\bar{C}_i$ coincide at leading logarithmic order with the Wilson
coefficients in the standard basis \cite{Buchalla:1996}. The
coefficients $C_i$ are calculated at the scale $\mu=m_W$, in a
perturbative expansion in powers of $\alpha_s(m_W)$, and are then
evolved down to scales $\mu \sim m_b$ using the renormalization
group equations.

The $\overline{\rm MS}$ masses $\overline{m}_b(\mu)$ and
$\overline{m}_s(\mu)$ are calculated according to
Refs.~\cite{Gray:1990, Chetyrkin:1999} and are given in
Table~\ref{tab:param2}.

\subsection{\label{subsec:form factors}
Form factors of $B \to K^*$ transition}

The hadronic part of the matrix element in Eq.~(\ref{eq:0A1})
describing the $B\to K^*e^+e^-$ transition can be parametrized in
terms of $B\to K^*$ form factors, which usually are defined as
\begin{equation}\label{eq:0A2}
\langle\bar{K}^*(k,\epsilon)|\bar{s} \gamma_\mu
b|\bar{B}(p)\rangle=\frac{2\,V(q^2)}{m_B+m_{K^*}}\,
\varepsilon_{\mu\nu\alpha\beta}\,\epsilon^{\nu*}p^{\alpha}k^{\beta}\,,
\end{equation}
\begin{eqnarray}
\langle\bar{K}^*(k,\epsilon)|\bar{s} \gamma_\mu\gamma_5
b|\bar{B}(p)\rangle&=&i\epsilon_\mu^*(m_B+m_{K^*})A_1(q^2)
\nonumber \\&&-i (\epsilon^*\cdot
p)(p+k)_\mu\frac{A_2(q^2)}{m_B+m_{K^*}} \nonumber
\\&&-i(\epsilon^*\cdot
p)\,q_\mu\frac{2\,m_{K^*}}{q^2}\nonumber
\\&&\times(A_3(q^2)-A_0(q^2)) \,,\label{eq:0A3}
\end{eqnarray}
with
\begin{eqnarray*}
A_3(q^2)&=&\frac{m_B+m_{K^*}}{2\,m_{K^*}}A_1(q^2)\\&&-
\frac{m_B-m_{K^*}}{2\,m_{K^*}}A_2(q^2)\,,\\ A_0(0)&=&A_3(0)\,;
\end{eqnarray*}
\begin{equation}\label{eq:0A4}
\langle\bar{K}^*(k,\epsilon)|\bar{s}\,\sigma_{\mu\nu} q^\nu
b|\bar{B}(p)\rangle=i\,2\,T_1(q^2)\,
\varepsilon_{\mu\nu\alpha\beta}\,\epsilon^{\nu*}p^{\alpha}k^{\beta}\,,
\end{equation}
\begin{eqnarray}
\langle\bar{K}^*(k,\epsilon)&|&\bar{s}\,\sigma_{\mu\nu}\gamma_5
q^\nu b|\bar{B}(p)\rangle\nonumber
\\&=&T_2(q^2)(\epsilon_\mu^*(P\cdot q) -(\epsilon^*\cdot
q)P_\mu)\nonumber
\\&&+T_3(q^2)(\epsilon^*\cdot q)(q_\mu-\frac{q^2}{P\cdot
q}P_\mu)\,,\label{eq:0A5}
\end{eqnarray}
with $T_1(0)=T_2(0)$. In the above equations, $q=p-k$, $P=p+k$,
$p^2=m_B^2$, $k^2=m_{K^*}^2$, $\epsilon^\mu$ is the polarization
vector of the $K^*$ meson, $\epsilon^*\cdot k=0$, and
$\varepsilon_{0123}=1$.

The $q^2$ dependence of the $B\to K^*$ form factors given in
\cite{Ball:2005} is parametrized as
\begin{equation}\label{eq:0A6}
F(q^2)=\frac{r_1}{1-q^2/m_R^2}+\frac{r_2}{1-q^2/m_{fit}^2}\,,
\end{equation}
\begin{equation}\label{eq:0A7}
F(q^2)=\frac{r_1}{1-q^2/m_{fit}^2}+\frac{r_2}{(1-q^2/m_{fit}^2)^2}\,,
\end{equation}
\begin{equation}\label{eq:0A8}
F(q^2)=\frac{r_2}{1-q^2/m_{fit}^2}\,,
\end{equation}
where the fit parameters $r_{1,2}$, $m_R^2$, and $m_{fit}^2$ are
shown in Table \ref{tab:ff2}.
\begin{table}[t]
\caption{The parameters $r_{1,2}$, $m_R^2$, and $m_{fit}^2$
describing the $q^2$ dependence of the $B\to K^*$ form factors in
the LCSR approach \cite{Ball:2005} and
$T_3(q^2)=\displaystyle\frac{m_B^2-m_{K^*}^2}{q^2}\left(\widetilde{T}_3(q^2)-
T_2(q^2)\right) $. The fit equations to be used are given in the
last column. } \label{tab:ff2}
\begin{center}
\begin{tabular}{c c c c c c}
\hline \hline & $r_1$ &$ r_2$ &$m_R^2\,,\rm GeV^2$ & $m_{fit}^2\,,\rm GeV^2$ & Fit eq.\\
\hline
$V$   & $0.923$ & $-0.511$ & $(5.32)^2$ &$49.40$ & (\ref{eq:0A6})\\
$A_1$ &         & $0.290$ &             & $40.38$ &(\ref{eq:0A8})\\
$A_2$ &$-0.084$ & $0.342$ &             & $52.00$ &(\ref{eq:0A7})\\
$A_0$ & $1.364$ & $-0.990$ & $(5.28)^2$ &$36.78$ & (\ref{eq:0A6})\\
$T_1$ & $0.823$ & $-0.491$ & $(5.32)^2$ &$46.31$ & (\ref{eq:0A6})\\
$T_2$ &         & $0.333$ &             & $41.41$ &(\ref{eq:0A8})\\
$\widetilde{T}_3$ &$-0.036$ & $0.368$ &           & $48.10$&(\ref{eq:0A7})\\
\hline \hline
\end{tabular}
\end{center}
 \end{table}

\section{\label{subsec:vector mesons} Amplitudes of $B \to K^* V$
decays }

An important ingredient of the resonant contribution is amplitude
of the decay of $B$ meson into two vector mesons, \ $B(p) \to
V_1(q,\epsilon_1) + V_2(k,\epsilon_2)$, with on-mass-shell meson
$V_2$ ($k^2 = m_2^2$) and off-mass-shell meson $V_1$ ($q^2 \ne
m_1^2$).

For the case of two on-mass-shell final mesons one can write the
amplitude in the form \cite{Valencia:1989}
\begin{eqnarray}
{\cal M} &=&\frac{G_F\,m_B^3}{\sqrt{2}}|V_{\rm CKM}|\Bigl(S_1\,
g_{\mu\nu} + \frac{S_2}{m_B^2}\, p_\mu p_\nu \nonumber \\&&- i
\frac{S_3}{m_B^2} \, \varepsilon_{\mu\nu\alpha\beta}\, q^\alpha
k^\beta \Bigr)\epsilon_1^{\mu *}\epsilon_2^{\nu *}\label{eq:0B1}
\end{eqnarray}
in terms of three invariant amplitudes $S_1$, $S_2$ and $S_3$,
$V_{\rm CKM}$ is a CKM factor. The quantities $S_1$, $S_2$ and
$S_3$ may be complex and involve two types of phases,
$CP$-conserving strong phases and $CP$-violating weak phases. In
general, the invariant amplitudes are a sum of several interfering
amplitudes, $S_{1j}$, $S_{2j}$ and $S_{3j}$, respectively. Then
the phase structure of $S_1$, $S_2$ and $S_3$ is:
\begin{equation}\label{eq:0B2}
S_k=\sum_{j}|S_{kj}|\,e^{i \varphi_{kj}}e^{i \delta_{kj}}\, \qquad
\qquad (k=1,2,3 ) \,,
\end{equation}
where $\varphi_{1j}$, $\varphi_{2j}$, and $\varphi_{3j}$ are the
$CP$-violating weak phases and $\delta_{1j}$, $\delta_{2j}$, and
$\delta_{3j}$ are the $CP$-conserving strong phases.

Using $CPT$ invariance, we can represent the matrix element for
the charge-conjugate decay $\bar B(p) \to \bar
V_1(q,\epsilon_1)\,\bar V_2(k,\epsilon_2)$ as
\begin{eqnarray}\label{eq:0B3}
\overline{{\cal M}}&=&\frac{G_F\,m_B^3}{\sqrt{2}}|V^*_{\rm
CKM}|\Bigl(\bar S_1\, g_{\mu\nu} + \frac{\bar S_2}{m_B^2}\, p_\mu
p_\nu \nonumber \\&&+ i \frac{\bar S_3}{m_B^2} \,
\varepsilon_{\mu\nu\alpha\beta}\, q^\alpha k^\beta
\Bigr)\epsilon_1^{\mu *}\epsilon_2^{\nu *}\,,
\end{eqnarray}
where $\bar S_1$, $\bar S_2$, and $\bar S_3$ can be derived from
$S_1$, $S_2$, and $S_3$ by reversing the sign of the
$CP$-violating phase. Note that if the $B\to V_1\,V_2$ decay is
invariant under the $CP$ symmetry, then $\bar S_1=S_1$, $\bar
S_2=S_2$, and $\bar S_3=S_3$. On the other hand, if all
$CP$-conserving phases of invariant amplitudes are equal to zero,
then $\bar S_1=S^*_1$, $\bar S_2=S^*_2$, and $\bar S_3=S^*_3$.

The helicity amplitudes in terms of three invariant amplitudes,
$S_1$, $S_2$, and $S_3$ are:
\begin{eqnarray}\label{eq:0B4}
H_\lambda&\equiv&\Bigl(S_1\, g_{\mu\nu} + \frac{S_2}{m_B^2}\,
p_\mu p_\nu \nonumber \\ &&- i \frac{S_3}{m_B^2} \,
\varepsilon_{\mu\nu\alpha\beta}\, q^\alpha k^\beta
\Bigr)\epsilon_1^{\mu *}(\lambda)\epsilon_2^{\nu *}(\lambda)\,.
\end{eqnarray}

From the decomposition Eq.~(\ref{eq:0B4}) one finds the following
relations between the helicity amplitudes and the invariant
amplitudes $S_1$, $S_2$, $S_3$:
\begin{eqnarray} \label{eq:0B5}
H_0 &=& - \frac1 {2 \hat{m}_1 \hat{m}_2} \Bigl(
(1-\hat{m}_1^2-\hat{m}_2^2)  S_1 \nonumber \\&&+ \frac{S_2}{2}
\lambda(1,\hat{m}_1^2,\hat{m}_2^2) \Bigr), \nonumber \\
H_\pm &=& S_1 \pm \frac{S_3}{2}
\sqrt{\lambda(1,\hat{m}_1^2,\hat{m}_2^2)} ,
\end{eqnarray}
with $\lambda(1,\hat{m}_1^2,\hat{m}_2^2) \equiv (1-\hat{m}_1^2)^2
- 2\hat{m}_2^2(1+\hat{m}_1^2) +\hat{m}_2^4$ and $\hat{m}_{1(2)}
\equiv m_{1(2)}/m_B$.

Note that the polarized decay amplitudes can be expressed in
several different but equivalent bases. For example, the helicity
amplitudes can be related to the spin amplitudes in the
transversity basis $\left(A_0\,,A_\|\,,A_\perp \right)$ defined in
terms of the linear polarization of the vector mesons via:
\begin{equation}\label{eq:0B6}
A_0 = H_0\,,\quad A_\parallel = \frac{H_+ + H_-}{\sqrt{2}}\,,
\quad A_\perp = \frac{H_+ - H_-}{\sqrt{2}}  \,,
\end{equation}
$A_0$, $A_\|$, $A_\perp$ are related to $S_1$, $S_2$ and $S_3$ of
Eq.~(\ref{eq:0B1}) via
\begin{eqnarray}\label{eq:0B7}
A_0 &=& -\frac1 {2 \hat{m}_1 \hat{m}_2} \Bigl(
(1-\hat{m}_1^2-\hat{m}_2^2)  S_1 \nonumber \\&&+ \frac{S_2}{2}
\lambda(1,\hat{m}_1^2,\hat{m}_2^2) \Bigr), \nonumber \\
A_\parallel &=& \sqrt{2} \, S_1\,,\quad A_\perp =
\sqrt{\frac{\lambda(1,\hat{m}_1^2,\hat{m}_2^2)}{2}}\, S_3\,.
\end{eqnarray}
The amplitude $\bar A_\lambda$ ($\lambda=0\,,\|\,,\perp$) are
related to the invariant amplitudes of the $\bar B \to \bar
V_1\,\bar V_2$ decay by the formulas
\begin{eqnarray}\label{eq:0B8}
\bar A_0 &=& - \frac1 {2 \hat{m}_1 \hat{m}_2} \Bigl(
(1-\hat{m}_1^2-\hat{m}_2^2) \,\bar S_1 \nonumber \\&&+ \frac{\bar
S_2}{2}\, \lambda(1,\hat{m}_1^2,\hat{m}_2^2) \Bigr)\,, \nonumber
\\ \bar A_\parallel &=& \sqrt{2} \, \bar S_1\,, \quad \bar
A_\perp =-
\sqrt{\frac{\lambda(1,\hat{m}_1^2,\hat{m}_2^2)}{2}}\,\bar S_3\,.
\end{eqnarray}
If the $B\to V_1\,V_2$ decay is invariant under $CP$
transformation, then $\bar A_0=A_0$, $\bar A_\|=A_\|$, and $\bar
A_\perp=-A_\perp$.

The decay width is expresses as follows:
\begin{eqnarray} \label{eq:0B9}
\Gamma(B \to V_1 V_2)&=&
\frac{m_B\,\sqrt{\lambda(1,\hat{m}_1^2,\hat{m}_2^2)}}{16\pi}
\left(\frac{G_F m_B^2}{\sqrt{2}}\,|V_{\rm CKM}|\right)^2 \nonumber
\\ &&\times\left( |A_0|^2 + |A_\||^2 + |A_\perp|^2 \right) \,.
\end{eqnarray}

The matrix element for the $B_d^0\to  K^{*0}\,V$ decay, where
$V=\rho^0\,,\omega\,, \phi\,,J/\psi(1S)\,,\psi(2S)\,,\ldots$
mesons, we can represent as
\begin{eqnarray}\label{eq:0B10}
{\cal M}
&=&\frac{G_F\,m_B^3}{\sqrt{2}}|V^*_{tb}\,V_{ts}|\Bigl(S^V_1\,
g_{\mu\nu} + \frac{S^V_2}{m_B^2}\, p_\mu p_\nu \nonumber \\ &&- i
\frac{S^V_3}{m_B^2} \, \varepsilon_{\mu\nu\alpha\beta}\, q^\alpha
k^\beta \Bigr)\epsilon_1^{\mu *}\epsilon_2^{\nu *}.
\end{eqnarray}
Next, we define the normalized amplitudes:
\begin{equation} \label{eq:0B11}
h_\lambda \equiv \frac{A_\lambda}{\sqrt{\sum_{\lambda^\prime}
|A_{\lambda^\prime}|^2}}\,,  \quad  \sum_\lambda |h_\lambda|^2 =
1\,  \quad (\lambda, \lambda^\prime = 0,
\parallel, \perp)\,.
\end{equation}
By putting $m_1 = m_V$, \ $m_2 = m_{K^*}$ and using
(\ref{eq:0B9}), (\ref{eq:0B11}) we obtain the relation between the
amplitudes $h_\lambda$ and $A_\lambda$ of the process under study
$B_d^0 \to K^{*0}\, V$ for any vector meson $V=\rho^0\,,\omega\,,
\phi\,,J/\psi(1S)\,,\psi(2S)\,,\ldots$:
\begin{eqnarray} \label{eq:0B12}
h_\lambda^V &=& \frac{G_F m_B^2}{4\sqrt{2}}|V^*_{tb}\,V_{ts}|
\sqrt{\frac{ m_B\,\tau_B}{\pi \, {\rm BR}(B_d^0 \to K^{*0}\, V) }
}\nonumber \\&&\times\lambda^{1/4}(1, \hat{m}_V^2,
\hat{m}_{K^*}^2)\, {A}_\lambda^V,
\end{eqnarray}
where ${\rm BR}(\ldots)$ is the branching ratio of $B_d^0 \to
K^{*0}\, V$ decay and $\tau_B$ is the lifetime of a $B$ meson.

Solving Eqs.~(\ref{eq:0B7}) we find the scalars $S_1\,,S_2$ and
$S_3$, and then extend the helicity amplitudes $A_\lambda^V$ off
the mass shell of the meson $V$, i.e. for $q^2 \ne m_V^2$. We
introduce the phases $ \delta_\lambda^V \equiv {\rm
arg}(h_\lambda^V)$, $ \delta_i^V \equiv {\rm arg}(S_i^V)$, where
$i=1\,,2\,,3$. Then we have
\begin{eqnarray} \label{eq:0B13}
|S_1^V|&=&\frac{|A^V_\||}{\sqrt{2}}\, ,\quad
 |S_3^V| = \sqrt{\frac{2}{\lambda(1,\hat{m}_V^2,\hat{m}_{K^*}^2)}}
\, |A_\perp^V|\,,
\nonumber \\
|S_2^V| & = &
\frac{\sqrt{2}}{\lambda(1,\hat{m}_V^2,\hat{m}_{K^*}^2)} \Bigl(
8\hat{m}_{K^*}^2 \hat{m}_V^2 |{A}_0^V|^2 \nonumber \\&&+
    \left(1-\hat{m}_V^2-\hat{m}_{K^*}^2\right)^2 |{A}_\parallel^V|^2
    \nonumber \\
   &&+
    4\sqrt{2}
    \hat{m}_{K^*}\hat{m}_V(1-\hat{m}_V^2-\hat{m}_{K^*}^2)\nonumber
    \\&&\times
    |{A}_0^V| |{A}_\parallel^V|\cos(\delta_\parallel^V - \delta_0^V)
    \Bigr)^{1/2} \,,
    \nonumber \\
\sin(\delta_2^V -\delta_0^V) &=& -
\frac{\sqrt{2}}{|S_2^V|\,\lambda(1,\hat{m}_V^2,\hat{m}_{K^*}^2)}\nonumber
\\&&\times
        (1-\hat{m}_V^2-\hat{m}_{K^*}^2)|{A}_\parallel^V|
        \sin(\delta_\parallel^V - \delta_0^V)\,,
     \nonumber \\
\cos(\delta_2^V -\delta_0^V) &=&  -
\frac{\sqrt{2}}{|S_2^V|\,\lambda(1,\hat{m}_V^2,\hat{m}_{K^*}^2)}\nonumber
\\&&\times
    \Bigl((1-\hat{m}_V^2-\hat{m}_{K^*}^2)|{A}_\parallel^V| \cos(\delta_\parallel^V -\delta_0^V)
    \nonumber \\&& +2\sqrt{2}\hat{m}_V \hat{m}_{K^*}
    |{A}_0^V|\Bigr)\,,\nonumber \\\delta_1^V&\equiv&\delta_\|^V \,(mod\, 2\pi) \, ,
 \quad
\delta_3^V\equiv \delta_\perp^V \,(mod\, 2\pi) .
\end{eqnarray}
\begin{table}[tbh]
\caption{Branching ratio \cite{PDG:2010}, and decay amplitudes for
${ B}_d^0\to {K}^{*0}\,\rho^0$ \cite{Chen:2006}, ${B}_d^0\to {
K}^{*0}\,\omega$ \cite{Chen:2006} and ${ B}_d^0\to {
K}^{*0}\,\phi$, ${ B}_d^0\to { K}^{*0}\,J/\psi$, ${ B}_d^0\to {
K}^{*0}\,\psi(2S)$ \cite{PDG:2010}.}
\begin{center}
\begin{tabular}{c c c c c c}
\hline \hline
 $V$ & $\rho^0$ &$\omega$&$\phi$&$J/\psi$&$\psi(2S)$ \\
\hline $10^6{\rm BR}({ B}_d^0\to { K}^{*0}\,V)$&$3.4$&$2.0$&$9.8$&$1330$&$610$\\
$|h_0^V|^2$       & $0.70$ & $0.75$ & $0.480$&$0.570$&$0.46$ \\
$|h_\perp^V|^2$   &$0.14$  & $0.12$ & $0.24$&$0.219$&$0.30$\\
$\delta_0^V$ (rad)& & &$2.82$& & \\
${\rm arg}(h_\|^V/h_0^V)$ (rad)&$1.17$  & $1.79$ & $2.40$&$-2.86$&$-2.8$\\
${\rm arg}(h_\perp^V/h_0^V)$ (rad)&$1.17$& $1.82$ & $2.39$&$3.01$&$2.8$ \\
$10^4|S_1^V|$ &$1.17$&$0.81$&$2.66$&$33.64$&$28.86$ \\
$10^4|S_2^V|$ &$2.65$&$1.67$&$5.20$&$42.49$&$52.65$ \\
$10^4|S_3^V|$ &$2.31$&$1.64$&$5.28$&$115.28$&$153.00$ \\
$\delta_1^V-\delta_0^V$ (rad)&$1.17$  & $1.79$ & $2.40$&$-2.86$&$-2.8$\\
$\delta_2^V-\delta_0^V$ (rad)&$-2.11$  & $-1.53$ & $-0.84$&$0.90$&$1.62$\\
$\delta_3^V-\delta_0^V$ (rad)&$1.17$& $1.82$ & $2.39$&$3.01$&$2.8$ \\
\hline \hline
\end{tabular}
\end{center}
 \label{tab:ampl}
\end{table}


\end{document}